\documentclass[10pt,a4paper]{article}
\usepackage[explicit]{titlesec}
\usepackage{hyphenat}
\usepackage{ragged2e}
\RaggedRight
\usepackage{float}
\usepackage[T1]{fontenc}
\usepackage{amsmath, amsthm}
\usepackage{graphicx}
\usepackage{epsfig}
\usepackage{subcaption}
\usepackage{multirow}
\usepackage{soul}
\setul{.6pt}{.4pt}
\usepackage{mhchem}
\usepackage{geometry}
\usepackage{fancyhdr}
\usepackage[labelfont={footnotesize,bf} , textfont=footnotesize]{caption}
\makeatletter
\renewcommand\tagform@[1]{\maketag@@@ {\ignorespaces {\footnotesize{\textbf{Equation}}} #1.\unskip \@@italiccorr }}
\makeatother
\titleformat{\section}
  {\normalfont}{\thesection}{1em}{\MakeUppercase{\textbf{#1}}}
\titlespacing\section{0pt}{0pt}{-10pt}
\titleformat{\subsection}
  {\normalfont}{\thesubsection}{1em}{\textit{#1}}
\titlespacing\subsection{0pt}{0pt}{-8pt}

\makeatletter
\newcommand\sixteen{\@setfontsize\sixteen{16pt}{6}}
\renewcommand{\maketitle}{\bgroup\setlength{\parindent}{0pt}
\begin{flushleft}
\vspace{-.375in}
\sixteen\bfseries \@title
\medskip
\end{flushleft}
\textit{\@author}
\egroup}
\makeatother

\usepackage[sort&compress]{natbib}
\makeatletter
\renewcommand\@biblabel[1]{\textbf{#1.}\hfill}
\makeatother

\bibpunct{}{}{,~}{s}{,}{,}
\title{Dilatational-Plasticity Opens a New Mechanistic Pathway for Macromolecular Transport Across Polymeric Interfaces Yielding Solid-State Bonding}
\author{Nikhil Padhye and Ajay Vallabh \\
}

\begin{document}

\maketitle

\begin{abstract}
Bonding between polymeric interfaces is encountered widely in several industrial applications. Many of these bonding processes 
rely on time-consuming and temperature-dependent classical mechanism of polymer interdiffusion via reptation in a melt state. Here, for the first time, we 
report a new mechanistic pathway for achieving solid-state polymer bonding by triggering rapid macromolecular acceleration 
through mechanical deformation. Large-scale molecular simulations reveal that active plastic deformation in 
glassy polymers, at temperatures well-below the bulk (and surface) glass transition temperatures T$_g^b$ (and 
T$_g^s$), is sufficient to cause segmental translations of the polymer chains that lead to interfacial interpenetrations, and formation of new entanglements. 
The underlying mechanistic basis for this new type of bonding is identified as enhanced molecular-scale dilatations (or densifications) in conjunction with 
accelerated molecular mobility during plastic deformation. The reported mechanistic insights open promising avenues for 
designing new bonding technologies or material systems, and transformation of the existing ones to achieve
quick and energetically less intensive bonding. 
\end{abstract}

\section*{Introduction}

Thermoplastics are a class of polymers that can be softened and melted by the application of heat, and can be processed either in 
the heat-softened state or in the liquid state by extrusion, injection molding, 3D printing, calendering, thermoforming, etc. 
The products derived from thermoplastics (or their composites) have now become an integral part of the modern society 
owing to several desirable properties such as flexibility, light weight, and high strength. 
Joining or bonding of polymeric interfaces is of paramount importance in plastics manufacturing, and plays 
a central in role in polymer-based product fabrication.
Since the earliest discovery of polymer interdiffusion in 1963 above the glass transition temperature 
\cite{voyutskii1963role}, it has been widely accepted that when two polymers are brought into contact at a temperature \textit{above} the glass transition temperature T$_g$, under 
moderate contact pressure, the polymer molecules interdiffuse and form entanglements across the interface over the experimental timescales to yield bonding 
\cite{kausch1989polymer,prager1981healing,jud1979load,Jud1981,brown1991adhesion,wool1981theory,wool1995polymer,klein1990interdiffusion}.
These studies found that the strength of the interface depended on healing time, temperature, contact pressure, molecular structure, and other chemical or physical characteristics of the polymer, the 
interdiffusion proceeded via reptation until the interface disappeared \cite{de1981formation}, and 
the  interfacial toughness $G_c$ and shear strength $\sigma_s$ followed a time-dependent scaling 
of $t^{1/2}$ and $t^{1/4}$, respectively. Molecular simulations have also confirmed these bonding 
trends \cite{ge2013molecular,ge2014healing,ge2014tensile}. All these studies emphasized temperatures above T$_g$ as an essential requirement for bonding.
At temperatures well-below T$_g$, bonding between glassy polymers, due to interdiffusion, cannot be detected on experimental timescales 
because the timescales for relaxation 
in the glassy state are extremely large \cite{colby2000dynamic, hutchinson1995physical,jerome1997dynamics}, and the system is effectively 
frozen concerning any cooperative segmental motions ($\alpha-$relaxation)\cite{angell2000relaxation} to 
yield entanglements across the interface. In spite of frozen mobilities at temperatures below T$_g$, researchers have 
reported polymer bonding below T$_g$ \cite{boiko1998strength,boiko2014chain}, by arguing that the 
surface of a glassy polymer has a lowered glass transition temperature (T$_g^s$) compared to that of the bulk (T$_g^b$), 
and a noticeable molecular-scale diffusivity exists on the free-surface that enables bonding on experimental timescales of hours.  

Recently Padhye et al. \cite{padhye2017new} have reported that macromolecular transport, and thereby polymer bonding, can be rapidly 
triggered across the polymer interfaces through mechanical deformation, at temperatures well-below T$_g$ (several 10s of degrees below 
T$_g$), at which naturally enhanced molecular mobility on the free surface of a polymer is nonexistent. 
They reported solid-state bonding of different polymer films (with varying molecular weights) in time on the order of a fraction of a second. 
The reported levels of bonding strengths correlated with gross plastic strain, and all the polymer films were bonded to similar strengths in the 
same interval of time. The mechanism of macromolecular transport in this new type of deformation induced bonding (DIB) is noted to be 
quite different 
from reptation-based interdiffusion in the melt state, since, unlike 
reptation-based interdiffusion, where diffusion coefficient shows a strong molecular weight dependence, in DIB different 
polymer films with different molecular weights were bonded in the same interval of time. Additionally, the bond strengths in DIB 
exhibited a non-monotonic correlation with the imposed plastic strain unlike reptation-based healing where interfacial strength grows monotonically with respect to time until interface disappears. The mechanistic origins of this new DIB phenomenon have remained 
elusive thus far, and reported in this letter.  

Using large-scale molecular simulations we confirm that compressive plastic deformation of polymer
samples, held in intimate contact, triggers requisite macromolecular acceleration and opens a mechanistic pathway through 
molecular-scale dilatations that facilitate chain
interpenetrations across interfaces to cause interfacial bonding. Solid-state glassy samples of sizes $66.5 \times 66.5 \times 66.5a^{3}$ ($a$ denoting bead size), at a temperature 
T$=0.3u_o/K_b$, were constructed using coarse-grained Kremer-Grest (K-G) model (well-known to phenomenologically represent linear amorphous polymers \cite{ge2013molecular,ge2014healing,ge2014tensile,hoy2006strain}), with T$_g$ $\approx$ 0.445u$_o$/K$_b$, by quenching the equilibrated molecular melts (See Methods). 
(Also see Supplementary Section 1 for detailed explanation for choosing coarse-grained bead spring model.)
Free surfaces of the two films were brought into contact at Z$=0$ plane, and subject to plane-strain active plastic deformation up to different levels of compressive plastic 
strains (5\%, 10\%, 15\%, 20\%, 25\%, and 30\%), Figures 1a-b. Also See Supplementary Video 1. NPT conditions were maintained during compression. To evaluate the strength of solid-state 
bonding between the polymer films, the newly bonded interfaces between the films were subject to uniaxial tensile test, Figure 1c, and the strengths of the 
bonded samples were characterized by the debonding stress-strain curves, Figure 1d. The area under the debonding stress strain curve was chosen 
as the quantitative measure for work of fracture (W$_f$), Figure $1e$. Corresponding to different levels of bonding strain, the debonding 
tensile tests exhibit a narrow elastic region with maximum elastic stress of around 3.8$\sigma_z a^3/u_o$, 
after which yielding and strain hardening is noted.
The debonding tests revealed events of chain pull-outs and scissions (See Supplementary Video 2), thereby providing the molecular evidence that active plastic deformation during compression was 
sufficient to cause enhanced molecular mobility and chain interpenetration across the interfaces to yield 
bonding below the bulk-T$_g$. 
The non-monotonic trends of W$_f$ with respect to plastic strain are found to be 
consistent with increasing macromolecular reorientation in the direction 
of principal stretches, which causes a lowering in number of entanglements across the interface at higher strains (See Supplementary Section 4). 
These trends are consistent with the earlier experimental reports \cite{padhye2017new}.
Glass transition temperatures of the free surface (T$^s_g$) and the interior bulk (T$^s_g$) of the polymer samples were estimated
to be T$^s_g=0.39 u_o/K_B$ and T$^b_g=0.445 u_o/K_B$, respectively, Figure 1$g$. (See Methods and Supplementary Section 2). 
The mobility at the free surfaces at 
T$=0.3 u_o/K_B$ is negligible compared to the mobility at T$^s_g=0.39u_o/K_B$, indicating that the free surface 
of the polymer is in glassy state at T$=0.3 u_o/K_B$, and any natural long-range mobility is nonexistent \cite{padhyemolecular}. 
During active plastic deformation (in 0--30\% deformation range) mean-square-displacements of polymer segments (averaged over time intervals of $50\tau$ over entire set of beads in thin layers 
within the bulk and at the interface), are enhanced by one order to two orders of magnitude compared to the undeformed sample at
T$=0.3 u_o/K_B$.  (See Supplementary Section 2.)

\begin{figure}[H]
   \hspace{-.5cm} \includegraphics[scale=.85]{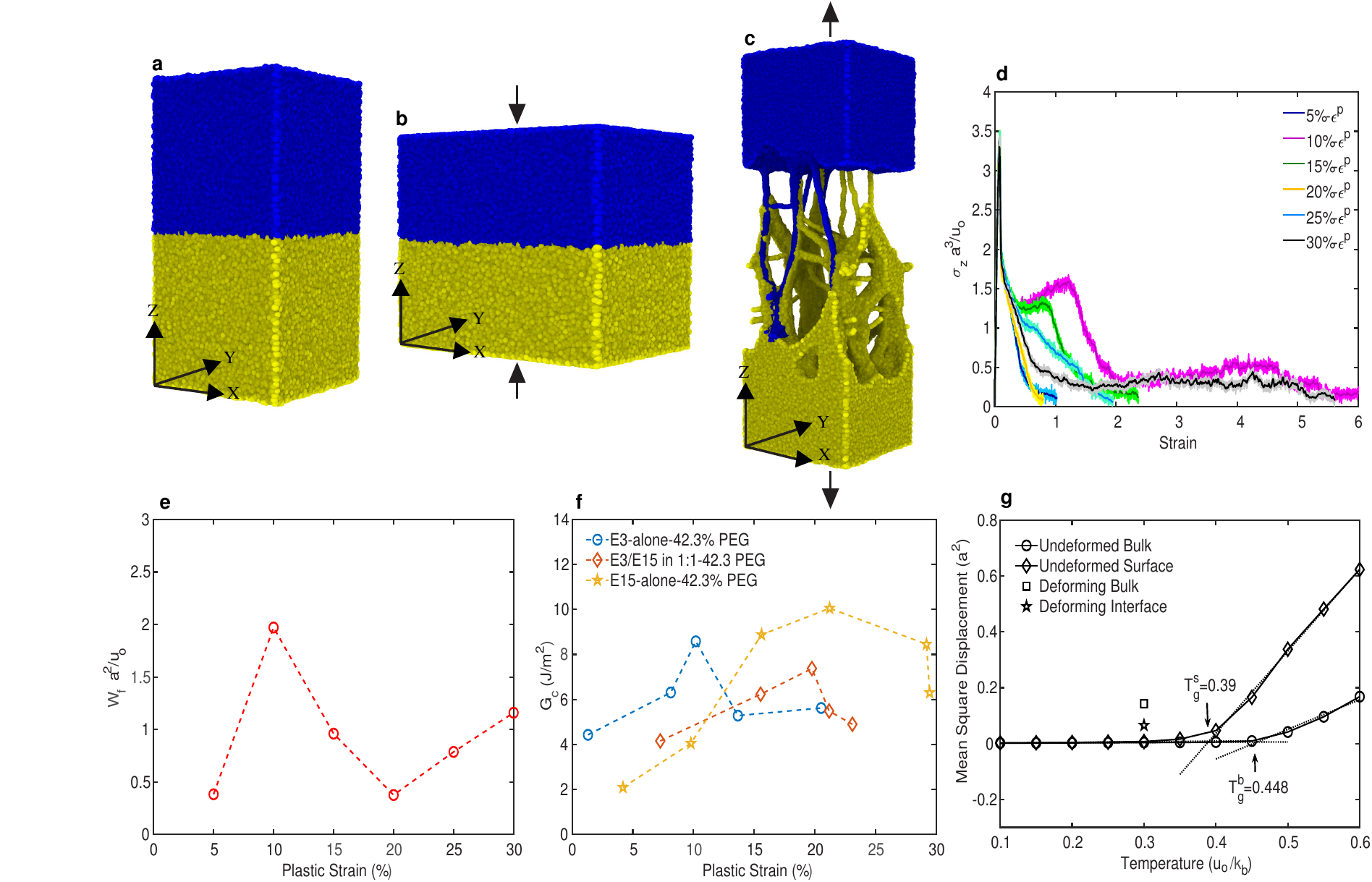}
    \caption{\textbf{Deformation-induced bonding of polymer samples under plane strain compression}.\textbf{a, b, c}, Represent states of two polymer samples before bonding, being subject to compressive plastic straining, and debonding during uniaxial tension, respectively. \textbf{d}, Debonding stress-strain response of bulk samples bonded at $5$\%, $10$\%, $15$\%, $20$\%, and $30$\% plastic strains. 
    $X\%$-$\epsilon^p$ abbreviates the imposed bonding plastic strain. The thin lines are raw stress-strain data, and the thick lines represent 
    smoothed data using a moving average filter.\textbf{e}, Debonding work of fracture $W_f$ (in $a^2/u_o$ units) versus plastic strain plot. 
    The $W_f$ values are based on the area under the stress-strain curves obtained from MD simulations. \textbf{f}, Fracture toughness $G_c (J/m^2)$ versus plastic strain plots for E3/E15 in 1:1-42.3\% polyethylene glycol (PEG), E3-alone-42.3\% polyethylene glycol (PEG), and E15-alone-42.3\% polyethylene glycol (PEG) taken from \cite{padhye2017new}. \textbf{g}, Estimation of surface and bulk glass transition temperatures $T_g^s$ and 
   $T_g^b$, respectively by calculating MSD as a function of temperature.}
    \label{fig:my_label}
\end{figure}

\begin{figure}[H]
    \centering
    \includegraphics[scale=0.97]{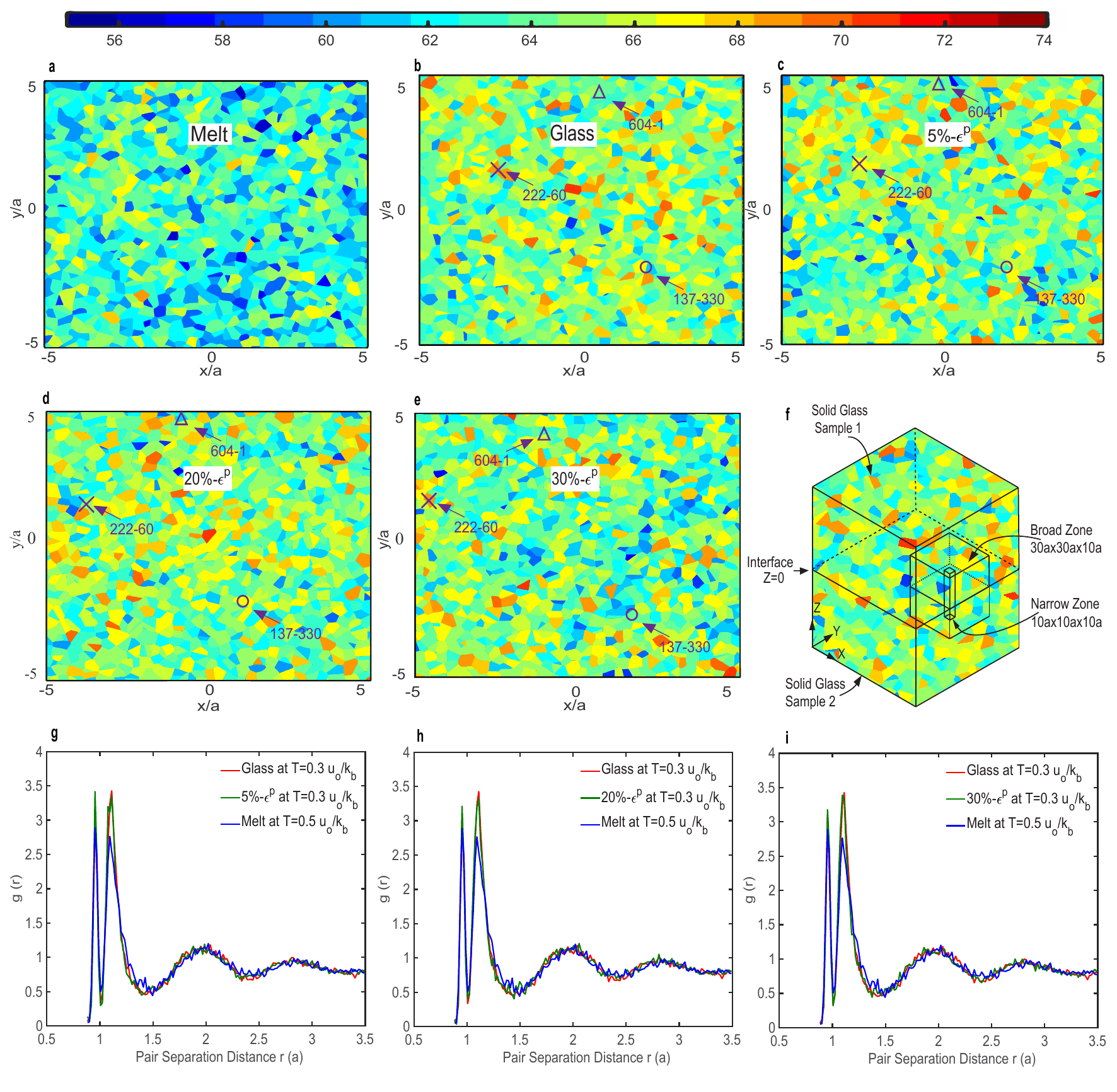}
    \caption{\textbf{Local number density and radial distribution function of small 
 10 ax10 ax10 a across the interface}. Local number density distribution in \textbf{a} polymer melt equilibrated $T$=$0.5u_o/K_b$, (b) quenched glass $T$=$0.3u_o/K_b$, and \textbf{c-e} samples deformed at $5$\%, $20$\% \& $30$\% 
 at $T$=$0.3u_o/K_b$, respectively, \textbf{f} region in which local number density distribution is calculated, 
  \textbf{g-i}, plots of radial distribution function (RDF) comparing DIB bonded samples at $5$\%, $20$\% \& $30$\%, with undeformed glass and the polymer melt.}
    \label{fig:my_label}
\end{figure}

To identify the mechanistic origins of DIB during plastic deformation, we investigate the evolution of microstructure by characterizing the changes in the local volume, segmental rearrangements, 
and molecular transport across the interfaces.  Previous studies \cite{deng1989simulation,argon1995plastic} have shown that the free volume in a 
glassy polymer is distributed over a wide range of atomic dimensions, and that this distribution is responsible for the relaxation phenomenon of viscoelasticity and diffuse shear transformations\cite{argon2013physics} which 
accelerate molecular mobility during deformation. Local liquid-like environments percolate through the glassy volume and enable the acceleration of segmental relaxations, 
and an actively deforming glass is identified as a heterogeneously dilated continuum with molecular mobility comparable to that at above T$_g$ \cite{zhou2001enhanced,argon1999mechanistic}.
We note that although macromolecular acceleration is a requirement for a polymer molecule to migrate across the interface and yield bonding, it is evidently not a sufficient condition to cause bonding because an accelerated
molecule near the interface can incrementally translate parallel to the bonding plane without causing any interpenetration. 

The classical notion of shear transformation zones (STZs), acting as carriers of plastic deformation, does not provide any molecular-scale 
information about the polymer dynamics during deformation, therefore
we define a local state variable called the local number density $\rho_N$ for each molecular site (coarse-grained bead). $\rho_N$ represents the 
number of beads present in a local spherical volume ($V_l$) of radius $r_l=2.5a$ centered at any reference bead. Since the LJ potential cut-off 
radius is $2.5a$, $\rho_N$ quantatively captures the local packing density or the liquidty factor at any molecular site. 
The average densities of the solid glass at T$=0.3 u_o/K_B$ and 
the polymer melt at T$=0.5u_o/K_B$ were found to be $\rho_g=1.003a^{-3}$ \& $\rho_l = 0.85a^{-3}$, respectively, and accordingly the average local number density for the glassy and the melt states are determined to be  $\rho_{N_g}=65.65$ and $\rho_{N_l}=55.63$, respectively.
Another metric called the radial distribution function (RDF) \cite{demkowicz2005autocatalytic,demkowicz2005liquidlike} is used to characterize the dynamics of the microstructure during deformation (See Methods and Supplementary Section 4). RDF provides 
the probability $g(r)$ of finding a bead that is at a distance $r$ from the reference bead, and captures the aggregated global information about the distribution of beads within the microstructure. 

Figures 2$a$-$e$ show the spatial distribution of $\rho_N$ in the equilibrated melt at T$=0.5u_o/K_b$, quenched glass at T=$0.3u_o/K_b$,  and glasses having undergone $5$\%, $20$\% \& $30$\% plastic deformation at T$=0.3u_o/K_b$, respectively, in a $10a\times10a \times10a$ sized narrow zone at the interface (Figure $2f$).
 Figures $2h$-$i$ show corresponding plots of RDFs for $5$\%, $20$\% \& $30$\% deformed samples compared with the RDFs of the 
quenched glass at $T$=$0.3u_o/K_b$ and the equilibrated melt at $T$=$0.5u_o/K_b$. The $\rho_N$ plot 
of the melt state (Figure 2$a$) is 
dominated by the ``blue regions'', indicating low local number density and loose molecular packing, a clear contrast compared to the glassy states 
(with or without deformation, Figures 2$b$-$e$). The quenched and deformed glasses (Figures 2$b$-$e$) are microstructurally heterogeneous, with local liquid-like 
packets embedded within the solid-state bulk. As deformation proceeds from 0\% to 30\% plastic strain, liquid-like regions appear and disappear at different molecular sites. 
As such there is no evidence of increased liquidity, i.e., appearance of more ``blue regions'', in the deforming glasses  compared to undeformed glass, but it is evidently clear that the local liquid-like regions are moving through 
the glassy bulk as the 
deformation proceeds. A particular molecular site which is densely packed at some instant can subsequently transform into a liquid-like region, and vice-versa, and these changes provide spatial allowance for molecular 
rearrangements. Since the global bulk deformation must follow kinetic compatibility due to imposed external loads at the boundary of the solids, the local liquid-like regions aid the cooperative relaxations of cluster of beads during incremental straining.  

The RDFs in Figures $2h$-$i$ reveal two primary sharp peaks followed by a diffused pattern. 
The first and second sharp peaks correspond to the distances ($r$) 
at which finite extensible nonlinear elastic (FENE) potential (the intra-particle pair potential) and the Lennard-Jones (LJ) potential  (the 
inter-particle pair potential) reach their minimum values, respectively.
Across all these plots, the RDF curves of various deformed glasses and the undeformed glass are similar, indicating that the distribution of the molecular distances, on an average, within the microstructure is not altered during deformation in any noticeable manner; whereas, for the melt state the first and second peaks are relatively wide and lowered, indicating a loosely packed behavior of fluid. Overall, we conclude that the plastic deformation in itself 
does not transform a solid-state glass into a melt state, 
and that the macroscopic behavior of glass during deformation at constant temperature is still solid, however, plastic deformation triggers heterogeneous evolution of microstructure, and  liquid-like regions of mobility within the glassy structure percolate through the bulk during deformation. The molecular dimension $2.5a$ emerges as the relevant length scale for capturing these local-structural changes.  

To understand the nature of evolution of molecular-scale behavior in deforming glasses, we followed the motion of a few randomly chosen beads, with IDs 604-1, 222-60, and 137-330 (See Supplementary Section 
for ID nomenclature), initially present in the ``narrow region'' within the glassy state (see marked molecular sites in Figure 2$b$), and found that, both, the local number density and positions of these 
beads changed noticeably as deformation proceeded (see the location of the marked beads in deformed glasses in Figs. 2$c$-$e$, and compare them with their locations in the undeformed glass). We also 
observed that the same beads in the glassy state, over deformation timescales (up to 1951$\tau$), in fact exhibited molecular-scale dilatations (lowering in local number density), and 
densifications (increase in local number density) at $T=0.3 u_o/k_B$; however, without any long-range displacements. The conformations of polymer chains were essentially frozen, and these dilatations or 
densifications within the glass, over the considered timescales, essentially corresponded to molecular-scale vibrations, which are non-zero even in solid-state at temperatures well-below the bulk or surface 
glass transition temperatures.

\begin{figure}[H]
    \centering
    \includegraphics[scale=0.95]{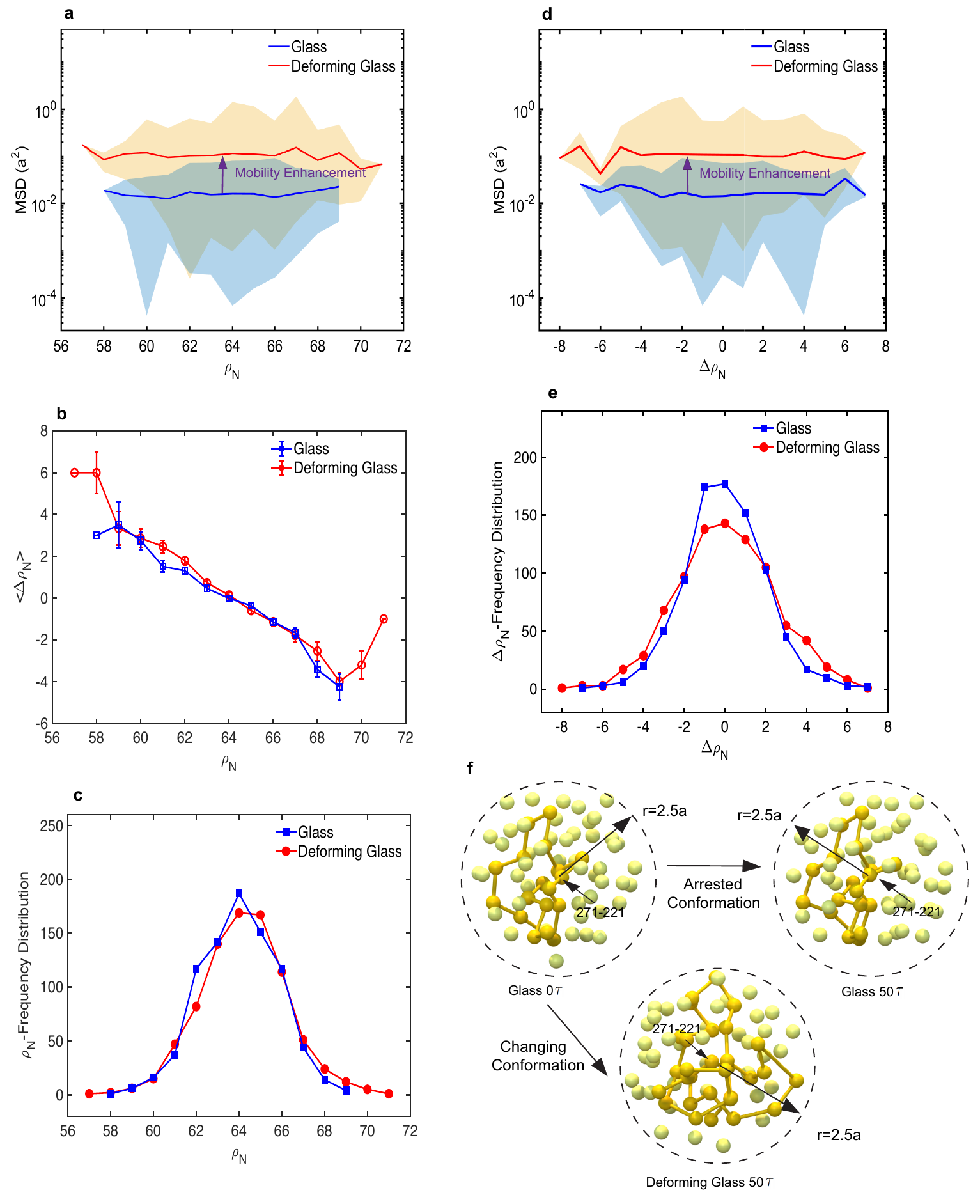}
    \caption{\textbf{Chain-End-Motion.} \textbf{a, d}, Comparison of chain-end displacements with respect to local number density, and gradient in local number density. \textbf{b}, correlation between gradient in local number density and local number density, \textbf{c} frequency distribution of local number density, \textbf{e} frequency distribution of gradient in local number density, 
    and \textbf{f} conformation arrest and transformation in the glassy state and deforming glass, respectively.} 
    \label{fig:my_label}
\end{figure}

In order to quantify and correlate the accelerated molecular mobility during plastic deformation with the changes in the microstructure, and contrast it with the arrested long-range 
mobility in glassy state, we computed 
the mean-square-displacements (MSD) 
of the chain-ends, both, in the glassy state and during the course of the deformation, and plotted them with respect to the time-dependent local number density ($\rho_N$), 
and temporal gradient of the local number density $\triangle \rho_N$, Figures 3$a$ and $d$. From these plots it is clear that there is at least an order to two orders of magnitude 
enhancement in the mobility in deforming glass compared to undeforming glass, for the entire range of $\rho_N$, and $\triangle \rho_N$. 
Most important characteristic feature of the MSD curves worth noting is that the quantitative values for MSD, in the deforming glass, are comparable to $a$ (bead size), implying the plausibility of 
sufficiently large molecular level openings through which beads of size $a$ can move across the interface and yield bonding. 
The plots of  
$\triangle \rho_N$ vs $\rho_N$, both, for the glassy state and deforming glass, are shown in Figure 3$b$. This correlation between $\rho_N$ and 
$\triangle \rho_N$ reveals that soft sites (low $\rho_N$ values) undergo densification (positive $\triangle \rho_N$), whereas as hard sites (high $\rho_N$ values) dilate (negative $\triangle \rho_N$).
Although occurrence of molecular-scale dilatations and densifications is noted, both, in undeformed and deforming glass, these volume changes 
are not accompanied by a similar level of enhanced mobility in the solid-state glass, when compared to the deforming glass. After observing the exhaustive data, 
we found that number of dilatations were roughly the same as the number of densifications (See Supplementary Section 4 for sample data), which is consistent with no noticeable changes 
in the RDF plots between deforming and undeformed glasses. The frequency distributions of $\triangle \rho$, both, for deforming and undeformed glass are shown in Figure 3$e$.
We find that glassy state is characterized by a higher number of non-volume changing events compared to the deforming glass, and in the deforming glass the number of events corresponding
to larger-size dilatations or densifications have increased. A similar behavior is noted for frequency distribution of $\rho_N$,  occurrence of relatively harder or softer sites 
(marked by extreme values of $\rho_N$) is more prominent in the deforming glass. Figure 3$f$ shows an example where, despite molecular-scale dilatations in the glassy state the 
conformation of polymer chains are essentially arrested, whereas during plastic deformation the conformations are noticeably altered, elucidating the role of deformation in 
enhancing mobility in conjunction with the molecular-scale volume changes.  Similar characteristic trends were found for non-chain-ends. 

The reptation dynamics of a polymer chain in an entangled melt is restricted by the topological constraints, and the motion of a polymer chain occurs along the contours of the tube in which it is trapped. Thus, during interdiffusion across interface, the chain-ends play a leading role in causing interpenetration and formation of new interfacial entanglements. In case of solid-state glassy networks, when deformation triggers enhanced mobility 
of the polymer segments, like in the polymer melts, the polymer chains cannot cross each other, nor can the polymer segments simply break their topological bonds and displace arbitrarily. 
Here, we have found the favorable molecular-scale dilatations or densifications, and enhanced mobility, provide 
opportunities for, both, the chain-ends  and the non-chain-ends  to interpenetrate across the interface to yield bonding, and in particular interpenetration of the chain-ends sometimes advances into formation of new entanglements. This is the key mechanistic difference between deformation-induced bonding and bonding between polymer melts. 
Figure 4$a$-$d$ illustrates how a polymer chain (ID 102), with a particular non-chain-end (ID 102-230) located near the interface at 0\% plastic 
strain, interpenetrated from the lower half into the upper half at 20\% plastic strain. Following spherical region of influence 
of radius $2.5a$, it is found that neighboring beads within this region are displaced by new neighboring beads upon incremental straining during the deformation period,
and the dilatation-plasticity under compressive traction causes cooperative relaxation of cluster of beads surrounding the non-chain-end, and the kinking-motion of the marked 
polymer segment enables relative slippage (or sliding) causing the non-chain-end
to penetrate into the other side of the interface (See Supplementary Video 3). This mechanism of molecular interpenetration is infeasible in an equilibrated molecular melt without deformation, as the polymer segments are constrained within their tubes. 
Another mode of interpenetration and subsequent entanglement formation is shown in Figures 4$e$-$h$. A chain-end (ID 504-1) of a polymer 
chain (ID 504), translates from the upper half into the lower half during molecular-scale volume fluctuations and enhanced mobility, accompanied by 
cooperative motion of the neighboring beads.
Even though disappearance of the interface and mixing of homopolymeric chains from opposite sides is 
favorable both energetically and entropically, we emphasize that once a glassy interface is subject to deformation, and the kinetically trapped glass is activated such that molecular-scale dilatations or densifications, and enhanced relative motions of chain segments ensue, it is not necessary that every bit of incremental deformation, and resulting segmental motion will necessarily cause interpenetration. The site of molecular-scale dilatation or densification, and incremental molecular motion must be compatible in a geometric sense to result in an interpenetration. The dependence of this interpenetration upon molecular parameters (composition, molecular weight, physical, and chemical properties), and processing conditions remains an open question at this stage, and requires further investigation.  
Finally, we illustrate the mechanism of chain-end interpenetration, followed by the formation of an entanglement, its disengagement, 
and reformation. Figures 4$i$-$l$ show primitive paths of two interfacial chains at 
$0$\%, $10$\%, $25$\%,\& $30$\% deformation, respectively. A primitive path \cite{edwards1977theory} is a geometrically constructed shortest path 
between the end-points of a reference chain in which the chain's contour can relax without crossing any obstacle. Each surrounding bead between 
the two ends of the chain along the primitive path represents a topological constraint, where the polymer chain forms entanglement with 
the surrounding chains. 
As the deformation proceeds, initially the randomly-coiled chain configurations are 
compressed, and the chain segments preferentially orient in the directions of principal stretches. In this process, whenever favorable 
molecular conditions emerge during dilatational-plasticity,  then the chain-end interpenetrates from the top-half into the bottom-half, and forms an entanglement. 
 We confirmed the formation of the entanglements using Z1-code \cite{hoy2009topological,karayiannis2009combined,kroger2005shortest} (see Methods for entanglement calculations). 
 Continued straining causes the disengagement of this entanglement, and 
subsequent reformation (See Supplementary Video 4). These behaviors explain the non-monotonic trends in deformation-induced bonding.

\begin{figure}[H]
    \centering
    \includegraphics[scale=0.95]{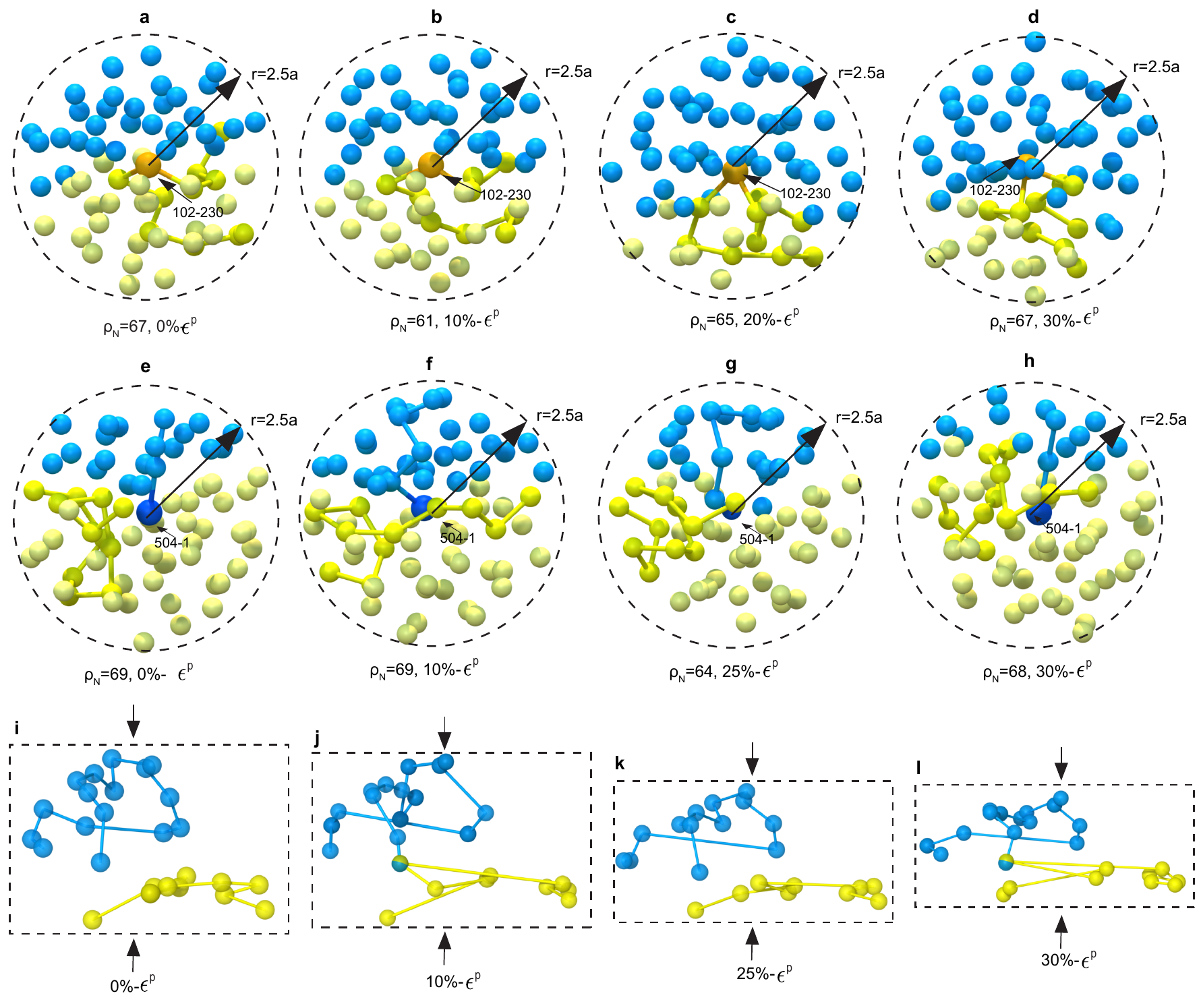}
    \caption{\RaggedRight\textbf{Molecular interpenetration, formation, disengagement, and reformation of entanglements.} 
    \textbf{a-d}, a non-chain-end with ID 102-230, at different stages of plastic strains 0\%-30\% with corresponding $\rho_N$, that exhibits interpenetration motion across the interface. 
    \textbf{e-h}, a chain-end with ID 504-1, at different stages of plastic strains 0\%-30\% with corresponding $\rho_N$, that also exhibits interpenetration across the interface. 
    \textbf{i-l} demonstrate primitive paths for opposite side chains at $0$, $10$, $25$, and $30$\% plastic strains, respectively. 
    Each blue and yellow beads represent topological constraints.}
    \label{fig:my_label}
\end{figure}

The results and discussions presented in this letter establish the mechanistic basis of deformation-induced bonding. The reptation timescales for the studied molecular system was estimated to be
$3M\tau$ \cite{ge2014tensile}, whereas the total deformation (0\%-30\%), and resulting bonding was achieved in 
timescales on the order of $1951\tau$. 
\section*{Methods}

\textbf{MD simulations.} Large-scale Atomic/Molecular Massively Parallel Simulator (LAMMPS) from Sandia National Laboratories was used to 
carry out all the molecular simulations. 
The bonding experiments were carried out in four key steps (a) equilibration of polymer melts, (b) formation of quenched 
glasses, (c) plastic compression of two polymer glasses held in molecular proximity, and (d) debonding tensile tests. See further details in Supplementary Section 1.  Kremer-Grest (KG) model was used to model polymer chains.  For LJ potential a 
cut-off radius $r_c=2.5a$ was chosen. For FENE the cut-off radius $R_o=1.5a$ and spring stiffness $k=30u_oa^{-2}$ were chosen. 
Each polymer sample was sized as $66.5\times66.5\times66.5a^{3}$ and 
contained $M=500$ polymer chains, with each comprising $N=500$ coarse-grained 
beads. The equilibration step was carried out under NPT conditions with $P=0$ at $T=1.0u_o/K_B$.  The equilibrated melt sample was 
quenched below to a solid-state glass at $T=0.3u_o/K_B$. Quenching rate of $\dot{T}=2 \times10^{-3} u_o/K_B\tau$ was used. Plane strain compressive plastic 
deformation was carried out between two polymer samples in the range of 0\% to 30\%. During debonding tensile test the two halves of the bonded sample 
were separated with a constant velocity $v=2\times0.005a\tau^{-1}$ leading to nominal strain-rate of $2 \times 10^{-4}\tau^{-1}$, and Quartic potential was used to model bond-breaking with parameter values of $K=2351u_o/K_b$, $B=-0.7425a$, $R_c=1.5a$, and $U_o=92.74467u_o$. All NVT and NPT simulations were 
performed using the Nose-Hoover thermostat and Nose-Hoover barostat. Newton's equation of motion was integrated using a velocity-Verlet 
algorithm with a time step $\delta t\leq0.01\tau$. 1 million $\tau$ is abbreviated as $1M\tau$.
\textbf{Calculation of surface and bulk glass transition temperatures.} The surface and bulk glass transition temperatures were calculated 
according to a standard procedure proposed in the literature \cite{morita2006study}. Starting from a temperature of $T=1.0u_o/K_b$ the system 
was cooled 
down to $0.1u_o/K_b$ at decrements of $0.05u_o/K_b$. At each temperature the system was first relaxed for $10000\tau$, and then  
simulated for $10000\tau$ under NPT conditions. During this period the mean square displacements ($MSDs$) were calculated in a thin layer, at the free surface and within the bulk, over intervals of $50\tau$. The inflection points on the MSD vs temperature plots were identified 
as the glass transition points. An alternate method of estimating glass transition temperature was also deployed, and gave similar results. (See 
Supplementary Section 2.)\\
\textbf{Calculation of $\triangle Z$ and $RMS$ to characterize molecular mobility.} The position vectors of the beads were extracted at 
increments of $0.5\tau$ during the 0-30\% deformation period. Corresponding values of $\rho_N$ and  $\triangle \rho_N$ were also calculated 
at these time intervals. The z-coordinate was utilized to compute incremental displacement $\triangle Z$ along the Z-axis, and used in 
computation of mean absolute z-displacement <|$\triangle Z$|>. Similarly, the incremental displacement vector $\triangle r$ was used to compute the RMS. 
See Supplementary Section 3 for complete details.\\
\textbf{Calculation of entanglements.} The motion of long chains is restricted by entanglements, which are topological constraints imposed by surrounding chains, entanglements were identified using the Z1 algorithm \cite{hoy2009topological,karayiannis2009combined,kroger2005shortest}. Z1 algorithm is based on a geometrical approach that determines entanglements directly without referring to statistics of the primitive path. This unique approach speeds up this algorithm several orders of magnitude compared to PPA \cite{everaers2004rheology}. See Supplementary Section 6. 
\section*{Supplementary Videos}

Video 1: Deformation-induced bonding between solid-state polymer samples.

Video 2: Debonding of the deformation-induced bonded interface.

video 3: Mechanism of bonding through motion of molecular kinks.

video 4: Formation of entanglement at the chain-end upon compression, followed by disengagement and reformation of the entanglement.

\section{Molecular Dynamics Simulations: Equilibration, Quenching, Bonding, and Debonding}
\label{md-simulations}
Coarse-grained (CG) models are computationally cheaper, capable of approaching larger spatial and temporal scales (compared to all-explicit atomistic models),  while retaining the 
essential chain-like nature of the molecules that cannot cross each other. CG-based molecular dynamics methods have been successfully applied in modeling the dynamics of diffusion in polymer melts \cite{padding2002time,kremer1990dynamics},
bonding through interdiffusion in polymer melts and debonding failure between polymer interfaces 
\cite{bulacu2009forced,bulacu2011effects,ge2013molecular,ge2014healing,ge2014tensile}, and mechanical behavior of glassy polymers (capturing the 
effects of strain rate, temperature, molecular entanglement density, and quenching rate \cite{hoy2006strain}). 
With the correct choice of intra- and intermolecular potentials, the coarse-grained models can produce mechanical response exhibiting softening, strain hardening, and anisotropy, 
which are critical features of a glassy polymer, and expected to affect the strength of deformation-induced bonded interface. Although coarse-grained models neglect the chemical details, scaling trends with respect to chain lengths, 
and intrinsic chain dynamics are expected to be universal in glassy polymers, therefore they are deemed as an appropriate tool for studying the phenomenological behavior of deformation-induced bonding. 

The polymer chains were represented using the Kremer-Grest  (KG) model. In the K-G model, 
monomers are represented as beads, and connected with each other through springs along the chain backbone. 
The beads interact with a truncated, and shifted Lenard-Jones (LJ) potential
$$
U_{LJ}(r)=4u_o\left[\left(\frac{a}{r}\right)^{12}-\left(\frac{a}{r}\right)^{6}-\left(\frac{a}{r_c}\right)^{12}+\left(\frac{a}{r_c}\right)^{6}\right],
$$

where $r_c$ is the cut-off radius, and was chosen as $2.5a$, such that, $U_{LJ}(r)=0$ for $r>r_c$, where $a$ is the bead size, u$_o$ is the binding energy,
and for the beads with mass $m$, the characteristic time is given as $\tau=a\left(m/u_o\right)^{\frac{1}{2}}$.  
In our LAMMPS simulations, all the quantities were expressed in LJ units. The interaction between topologically connected beads 
of a single chain are commonly modeled using an unbreakable finitely extensible non-linear elastic (FENE) potential

$$
U_{FENE}(r)=-\frac{1}{2}kR_o^2ln\left[1-\left(\frac{r}{R_o}\right)^{2}\right].
$$

For FENE, $R_o$ (equilibrium bond length for beads), and $k$ (bond coefficient for beads)
were chosen as $1.5a$, and $30u_oa^{-2}$, respectively. LJ and FENE potentials were used during equilibration, quenching and deformation-induced 
bonding simulations. During debonding, in uniaxial testing, the FENE potential was replaced with a Quartic potential \cite{stevens2001interfacial}, 
$$
U_Q(r)=K(r-R_c)^{2}(r-R_c)(r-R_c-B)+U_o,
$$
which allows for bond-breaking
between the beads as the chains are stretched. The Quartic potential
parameters were chosen as $K=1434.3u_o/K_b$, $B=-0.7589a$, $R_c=1.5a$, and $U_o=67.2234u_o$.
These parameters ensured that the equilibrium bond length based on $U_Q(r)$ was same as that from $U_{FENE}$.

\begin{figure}[H]
    \centering
    \includegraphics[scale=0.5]{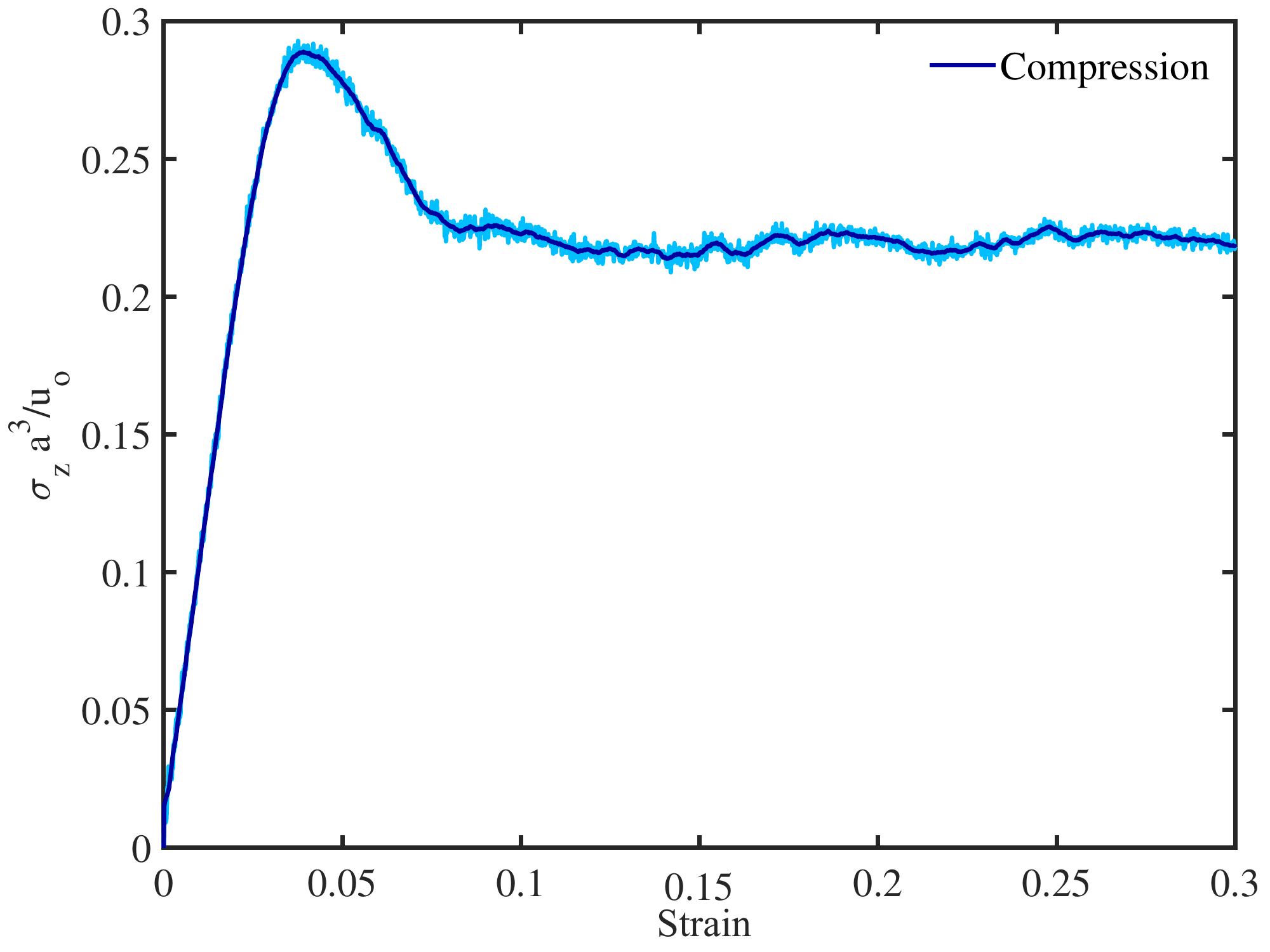}
    \caption{Stress-strain response corresponding to compression bonding between two polymer samples. }
    \label{fig:compression-test}
\end{figure}

Polymer samples of size $66.5\times66.5\times66.5a^{3}$ were constructed using a standard method from the literature \cite{auhl2003equilibration}. 
Each sample comprised $M=500$ polymer chains, with each chain containing $N=500$ beads. 
Periodic boundary conditions were used in all directions during the initial sample generation. The chains were then unwrapped in the $Z$-direction, 
rigid walls were imposed on top and bottom surfaces, and chains were compressed to produce the desired size initial box. 
This sample was equilibrated at $T=1.0 u_o/k_B$, under NPT conditions, with fixed boundary conditions in the $Z$-direction (imposed by  
repulsive confining walls). The sample pressure was maintained at $P=0$ by allowing expansion or contraction along $X$-direction,
with damping pressure coefficient $Pdamp=1000*dt$. 
The equilibrated melt sample was quenched to a temperature $T=0.3u_o/k_B$, 
at a quenching rate of $\dot{T}=2\times10^{-3} u_o/K_B\tau$, below the glass transition temperature 
(which was estimated to be $T_g\approx0.448u_o/K_b$), to produce
a solid-state glass. Pressure $P=0$ was maintained, and NPT ensemble, with damping pressure coefficient $Pdamp=1000*dt$ for $P_{xx}$ and $P_{yy}$, was used.
The quenched sample was re-equilibriated (for $0.1M\tau$) to eliminate any residual stresses. Density of the quenched glass was found to be $\rho\approx1.003$$a^{-3}$ (consistent with what is expected in a 
solid-state glass).
To perform bonding experiments, the free surfaces of two glass samples were first brought into molecular proximity without overlap at $Z=0$.
The repulsive walls, introduced during equilibration, were removed at the common interface such that free surfaces came into molecular proximity under 
Van der Waals forces. The two glasses were then compressed by moving the top and bottom walls along the Z-axis in opposite directions. Plane strain conditions were maintained in $Y$-direction,
and due to compression in the $Z$-direction the sample expanded along the $X$-direction.
Once the sample assembly was deformed to a desired level of compression, it was equilibrated for $0.1M\tau$ to ensure elastic recovery.
Different bonded samples, corresponding to different levels of imposed plastic strain, were prepared. Figure ~\ref{fig:compression-test} shows a sample stress-strain curve during compression bonding, which is consistent with solid-state glassy behavior.  

Debonding experiments were performed by subjecting the bonded sample  to uniaxial tensile loading. A central region of height $L_z^{o}=50a$, centered around 
the bonded interface, in the bonded assembly was selected for the debonding tensile test. Material regions outside this central region were rigidly held and displaced in 
opposite directions at a constant velocity $v=0.005\tau^{-1}$ 
along the $Z$-axis. The corresponding stress-strain curves, based on true stress and nominal strain measures, were collected. 
The tensile tests were carried under NVT conditions, and the temperature was maintained at $T=0.3u_o/k_B$, by using a damping temperature rate
 $Tdamp=100*dt$.
 
\section{Calculation of Surface and Bulk Glass Transition Temperatures}

To compute glass transition temperatures, we followed two methods: (a) free volume approach, and (b) MSD (mean square displacements) approach.  
The free volume approach is a standard procedure for computing the glass transition temperature of the polymer bulk,
whereas the RMS approach has emerged as an effective method for computing the glass transition temperature on the polymer free surface \cite{morita2006study}; however, bulk glass transition temperatures can also be computed 
using MSD approach. We compared the bulk glass transition temperatures according to both the approaches, and found an excellent match. 
In both these approaches, a melt sample of size $66.5\times66.5\times66.5a^{3}$ was prepared according to the procedure 
described in Section ~\ref{md-simulations}, 
with periodic boundary conditions in all directions. The sample comprised $M=500$ polymer chains, with each chain containing $N=500$ beads,
and parametric values for LJ and FENE potential chosen same as listed in Section ~\ref{md-simulations}.
The sample was equilibrated at a temperature $T=1.0u_o/k_B$, and pressure $P=0$. 
Further details for both these approaches are given as follows. 

\begin{figure}[H]
    \centering
    \includegraphics[scale=0.5]{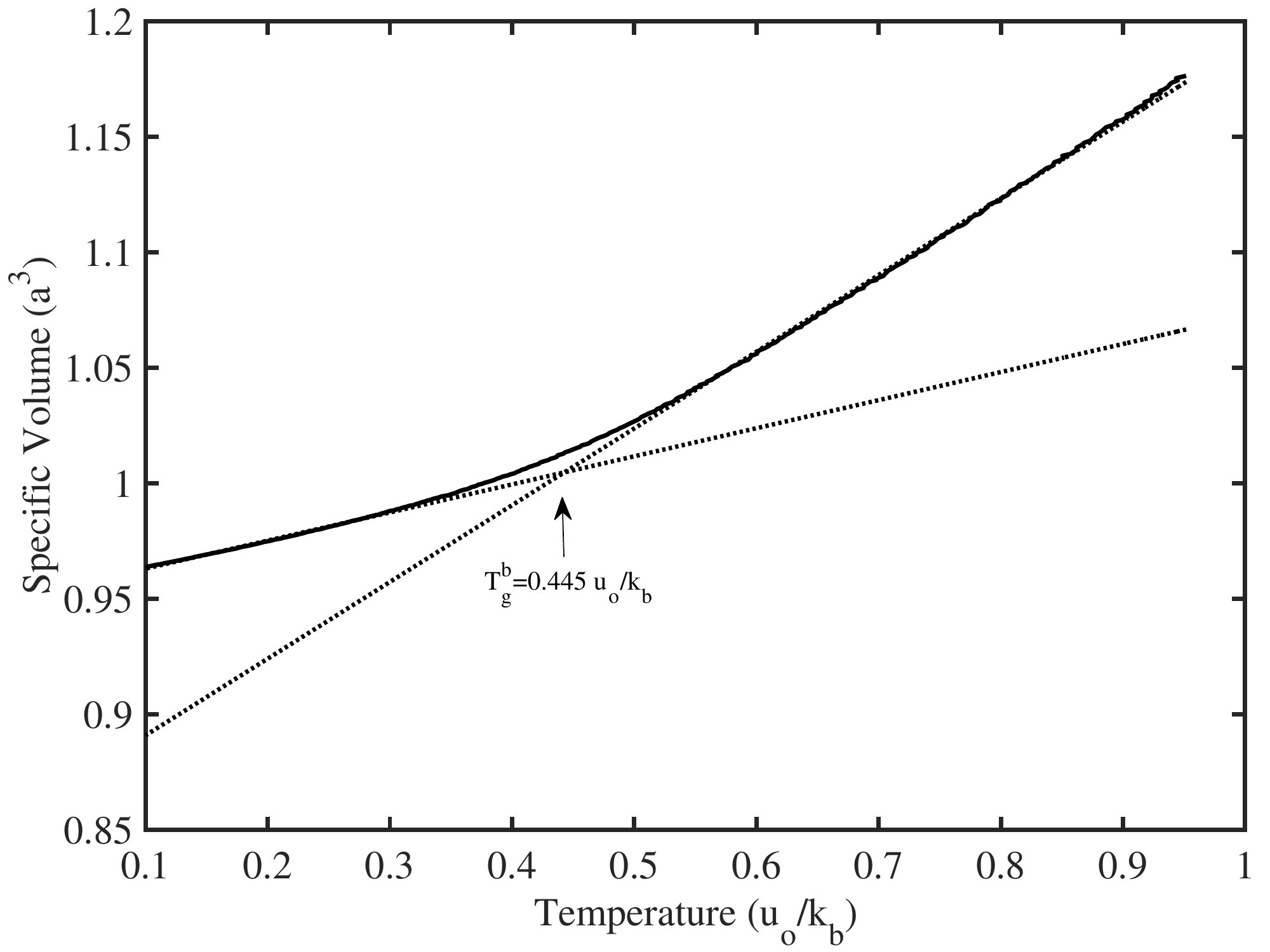}
    \caption{Free volume approach to compute the bulk glass transition temperature ${T_g}^s$.  ${T_g}^s$ is estimated to be $0.445$$k_B/\tau$. }
    \label{fig:compute-tg-free-volume}
\end{figure}

\textbf{(a) Free volume approach:} After a well-equilibrated melt sample was prepared, it was quenched from a temperature $T=1.0u_o/k_B$ to $T=0.1u_o/k_B$, at a quenching rate of $\dot{T}=2\times10^{-3}u_o/(k_B\tau)$, and constant pressure $P=0$. An NPT ensemble with pressure damping coefficient $Pdamp=1000*dt$ was used during quenching. The sample was allowed to expand and contract in all directions. Specific volume 
(averaged over the entire sample) with respect to temperature was collected at 
an interval of $1\tau$ during quenching. Figure ~\ref{fig:compute-tg-free-volume} shows the specific volume vs temperature plot. The inflection point on this plot gave an estimate for the bulk glass transition temperature as $0.445$$k_B/\tau$. 

\textbf{(b) RMS approach:} In this approach, the glass transition temperatures at the surface and the bulk were determined by evaluating the mean square displacements (MSD) of the polymer segments in the surface and bulk regions, respectively, as a function of temperature. We first started with the same well-equilibrated melt, as described earlier, and then the sample was quenched in multiple stages (quasi-static manner) 
from a temperature $T=1.0u_o/k_B$ to $T=0.1u_o/k_B$, 
at a quenching rate $\dot{T}=2\times10^{-3}u_o/(k_B\tau)$, and constant pressure $P=0$. An NPT ensemble with damping temperature and pressure 
coefficients was chosen as $Tdamp=100*dt$, and $Pdamp=1000*dt$, respectively.  During quenching the sample was allowed to expand and contract in $X$- 
and $Y$-directions. At every $0.05$$u_o/k_B$ unit reduction in temperature, the system was relaxed for $10000\tau$, and then again simulated for $10000\tau$ 
for collecting the mobility data. To collect the mobility data, we divided the polymer sample into multiple layers along the $Z$-axis, with each layer of thickness 
$a$ units. Three layers were chosen, both, at the surface and in the bulk, and position vectors of the polymer segments in these layers were documented
at increments of $50\tau$. Accordingly, mean square displacements, averaged over all segments in the respective surface and bulk layers, were computed. The 
inflexion point in the MSD vs Temperature plot, both, in surface and bulk layers led to the estimation of surface and bulk glass transition temperatures, $T_g^s$ 
and $T_g^b$, respectively. 

The quantitative data for MSD at different temperatures in the bulk and surface layers is given in the first two columns of 
Table ~\ref{tab:table-mobility}. Clearly, the MSD decreases as the temperature is decreased. We also computed the mobility data for the deforming glass at $T=0.3u_o/K_B$, 
over an interval of 50$\tau$, and found MSD to be $0.14251$$a^2$ and $0.06523$$a^2$ in bulk and interfacial layers, respectively. The MSD values in undeforming glass at $T=0.3u_o/K_B$ in bulk and surface layers were found to be $0.00368$$a^2$ and $0.00699$$a^2$, respectively. These data show two orders and one order enhancement in 
molecular mobility in bulk and surface regions, respectively, in a deforming glass compared to the undeformed glass.

\begin{table}[htb]
    \centering
    \caption{The MSD data for estimating bulk and surface glass transition temperatures. Data is listed for MSD in deforming glass.}
    \begin{tabular}{|l|l|l|l|}
    \hline
    \textbf{Temperature} & \textbf{MSD in the}  	  & \textbf{MSD in the} 	 & \textbf{MSD during $0$-$30$\% } 		          \\ 
    		$(u_o/k_B)$	 & \textbf{bulk layer $(a^2)$}     & \textbf{surface layer $(a^2)$}& \textbf{deformation at} T = 0.3$u_o/K_B$								    \\ \hline
 $0.1$   		  & $0.00174$		        & $0.00237$         & \multirow{8}{*}{\parbox{5cm}{Bulk layer: $0.14251$ ($a^2$)\\Interfacial layer: $0.06523$ ($a^2$)}} \\    
$0.15$ 				 &  $0.00189$ 		 & $0.00285$		        &								     	    									\\    
    $0.2$				 & $0.00245$ 		 & $0.00348$ 		 & 								           									\\    
    $0.25$ 			 & $0.00299$ 		 & $0.00521$ 		 &	   		           									\\    
    $0.3$ 				 & $0.00368$ 	 	 & $0.00699$ 		 & 			  						\\    
$0.35$ 				 & $0.00419$ 	        & $0.01576$ 	        &									         		  						 \\ 
$T_g^s\approx0.4$      & $0.00520$                & $0.04623$ 		 & 								        		 							 \\ 
$T_g^b\approx0.45$    & $0.00902$ 		 & $0.16540$ 		 &							                     								 \\     \hline
    \end{tabular}
    \label{tab:table-mobility}
\end{table}

\section{Computation of <|$\Delta z$|>, <MSD>, $\rho$, and $\Delta \rho$}

To understand how the local environment of a bead affects its mobility, we defined four time-dependent quantities: 
mean absolute Z-displacement (<|$\Delta z$|>), mean square displacements (<MSD>, different from the manner in which MSD was defined in the 
previous section),  local number density $\rho_N$, and 
local number density gradient $\Delta \rho_N$. These quantities can be calculated for any CG bead at any instant. 

\begin{figure}[H]
    \centering
    \includegraphics[scale=1.0]{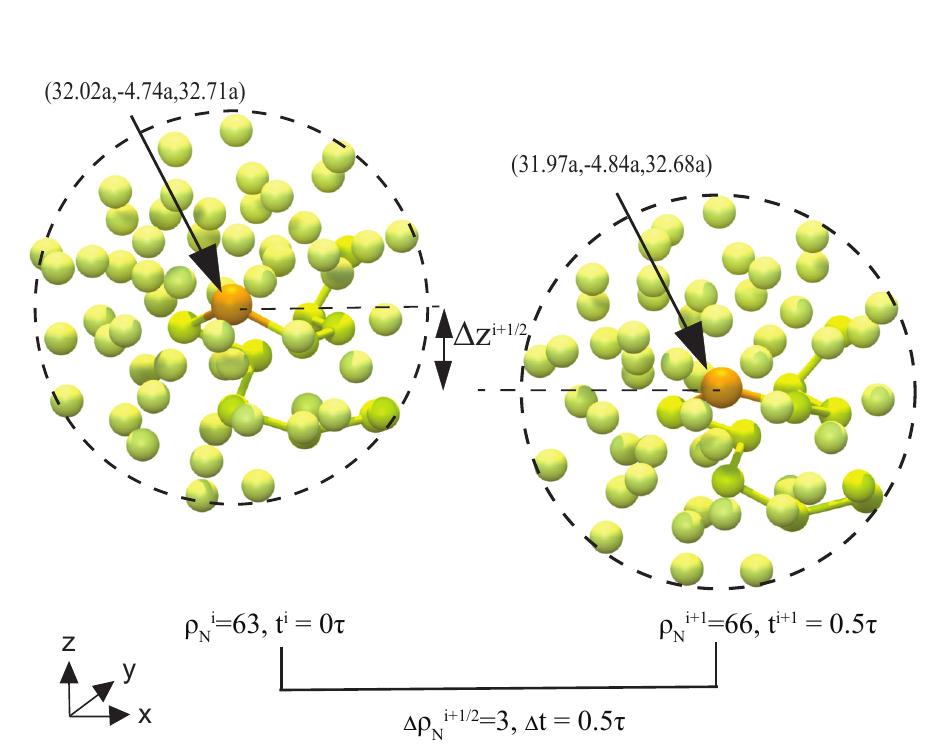}
    \caption{Illustration for computing local number density gradient and associated Z-displacement.}
    \label{fig:timed-data}
\end{figure}

Consider a discrete time-dependent motion where a bead at two subsequent time steps 
($t^i$) and ($t^{i+1}$) has position vectors, $Z$-coordinates, and local number densities as $\overrightarrow{r}^{i}$ and $\overrightarrow{r}^{i+1}$,
$z^i$ and $z^{i+1}$, and $\rho^i_N$ and $\rho^{i+1}_N$, respectively. We then define following quantities: 
gradient in local number density ${\Delta \rho_N}^{i+\frac{1}{2}} = {\rho_N}^{i+1}-{\rho_N}^{i}$,
square Z-displacement $|\Delta z^{i+\frac{1}{2}}|^2 = (z^{i+1}- z^{i})\cdot(z^{i+1}- z^{i}$), and square displacement 
$|\Delta \overrightarrow{r}^{i+\frac{1}{2}}|^2 = (\overrightarrow{r}^{i+1} - \overrightarrow{r}^{i})\cdot(\overrightarrow{r}^{i+1} - \overrightarrow{r}^{i})$.
For any bead, at any given time increment ($t^{i+1}-t^{i}$), we associate incremental mobility characterized 
in terms of $|\Delta z^{i+\frac{1}{2}}|^2$, and $|\Delta \overrightarrow{r}^{i+\frac{1}{2}}|^2$ with ${\Delta \rho_N}^{i+\frac{1}{2}}$, 
and $\rho_N^i$. An example of computing $Z$-displacement and gradient in the local number density is shown in Figure ~\ref{fig:timed-data}. 

Now depending on the choice of incremental time step, considered beads, and duration of the motion, 
for each value of $\rho_N$ and $\Delta \rho_N$ we can compute the associated 
mean square Z-displacement <$|\Delta z^{i+\frac{1}{2}}|^2$>, and mean square displacement <$|\Delta r^{i+\frac{1}{2}}|^2$>, by taking simple 
averages. The correlations among $\rho_N$, $\Delta \rho_N$, <$|\Delta z^{i+\frac{1}{2}}|^2$>, and 
<$|\Delta r^{i+\frac{1}{2}}|^2$> provide insightful information regarding deformation mechanics of polymer glasses, and its effect on bonding. We collected 
molecular states from LAMMPS at every $0.5\tau$, and then  
found that a time interval of $50\tau$ was an appropriate period to capture local relative motions between polymer segments. (At $0.5\tau$ interval resolution we found that only local bead vibrations were being captured.)

\section{Dilatational plasticity}

In this section, additional information and results are provided on computation of Radial Distribution Function (RDF), 
nature of plasticity-mediated enhanced mobility, distribution of local number density, growth of interface, entanglement formation, and 
results to enhance the performance of deformation-induced bonding.  

\begin{figure}
\centering
    \includegraphics[scale=0.75]{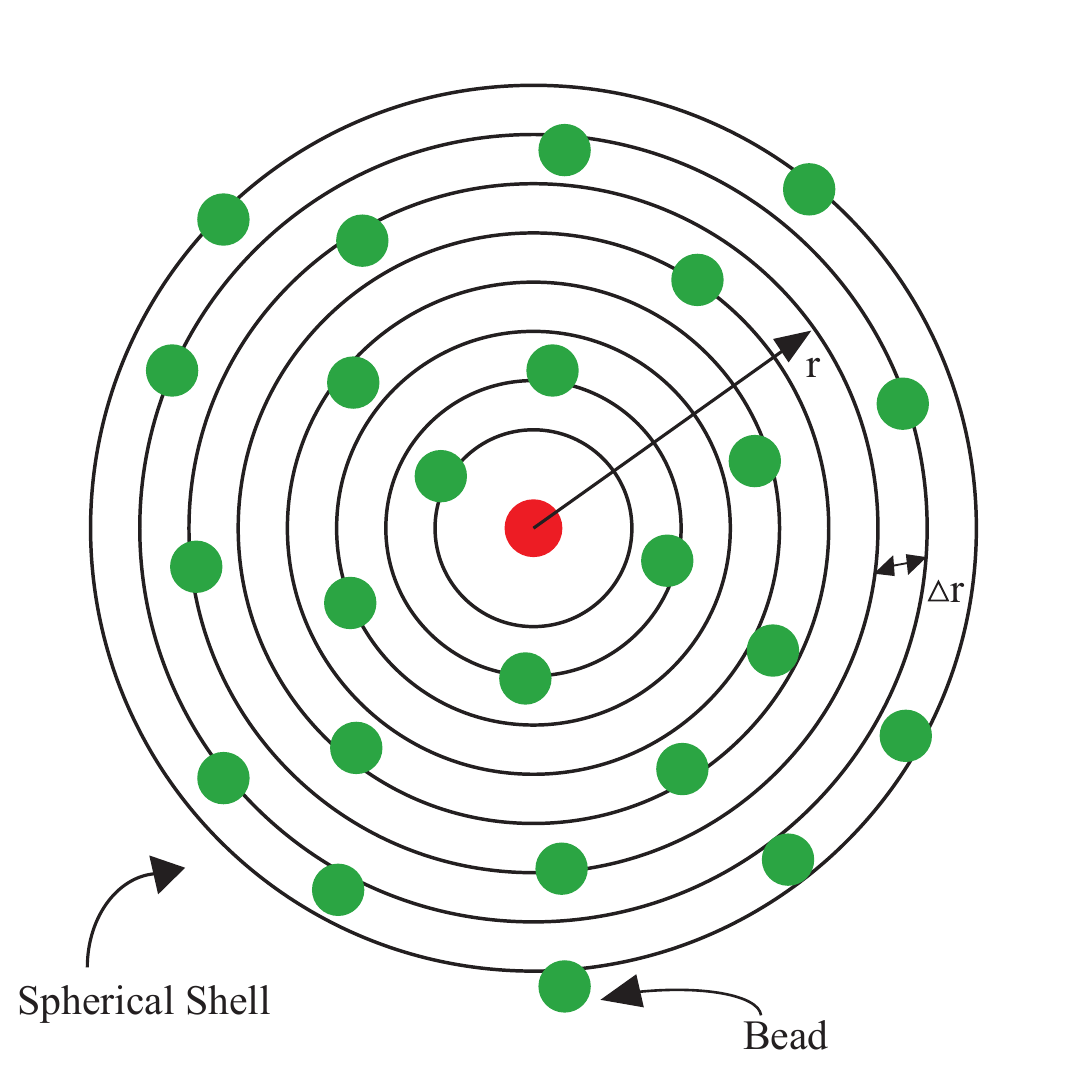}
    \caption{Schematic explaining the calculation of Radial Distribution Function (RDF).}
        \label{RDF-calculation}
\end{figure}

\textbf{Calculation of Radial Distribution Function (RDF):} The RDF represents the probability $g(r)$ of finding a bead in a shell that is at a distance $r$ from a reference bead. It also captures how closely are the beads packed together in a molecular system. To construct RDF, for any bead in the system, 
a series of concentric spherical shells are drawn, separated by a fixed distance $\Delta r$ (see Figure ~\ref{RDF-calculation}), and average number of beads 
($n(r)$) in an annular shell region, at distance $r$ and thickness $\Delta r$, are computed. Then, to obtain $g(r)$, $n(r)$ is divided by the volume of each annular 
shell ($4\pi r^2\Delta r$), and the average particle density ($\rho_{avg}=N/V_{box}$, where $N$ is the total number of particles in the system, and $V$ is total 
volume of system).  The RDF is formally given as:                            
$$
g(r)=n(r)/(\rho_{avg}*4\pi r^2\Delta r).
$$

In the present analysis, $\Delta r$ is chosen as $r_{cut-off}/h$,  where $r_{cut-off}$=$3.5a$, and $h=200$. LAMMPS simulations provide molecular states which are used to construct RDF plots in the commercial post-processing software OVITO. 

Table ~\ref{tab:dilatation-densification-data} shows a summary of dilatation and densification data collected for some chain-ends in the interfacial region 
during deformation.
The chain-ends are labeled according the ID of the chain to which they belong, followed by their position in the chain, e.g., 
chain-end ID $29$-$1$ indicates that this chain-end belongs to the chain with ID $29$, and it has a position number 1 in the chain. 
From the tabulated data, we conclude that number of dilatation events are approximately same as number of densification events, which is 
consistent with the fact that during deformation RDF plots in deforming glasses are not altered in any noticeable manner. 

\begin{figure}
    \begin{subfigure}{0.45\linewidth}
    \caption{\RaggedRight\textbf{}}
    \includegraphics[scale=0.5]{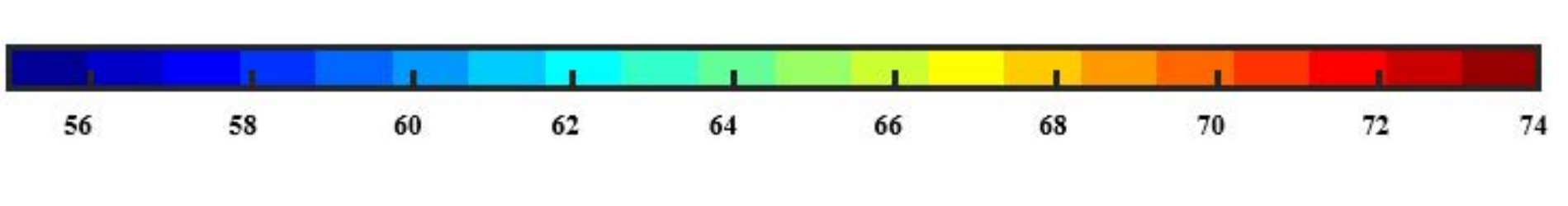}
        \label{fig:Colorbars.png}
        \end{subfigure}
        \par
    \begin{subfigure}{0.45\linewidth}
    \caption{\RaggedRight\textbf{}}
    \includegraphics[scale=0.24]{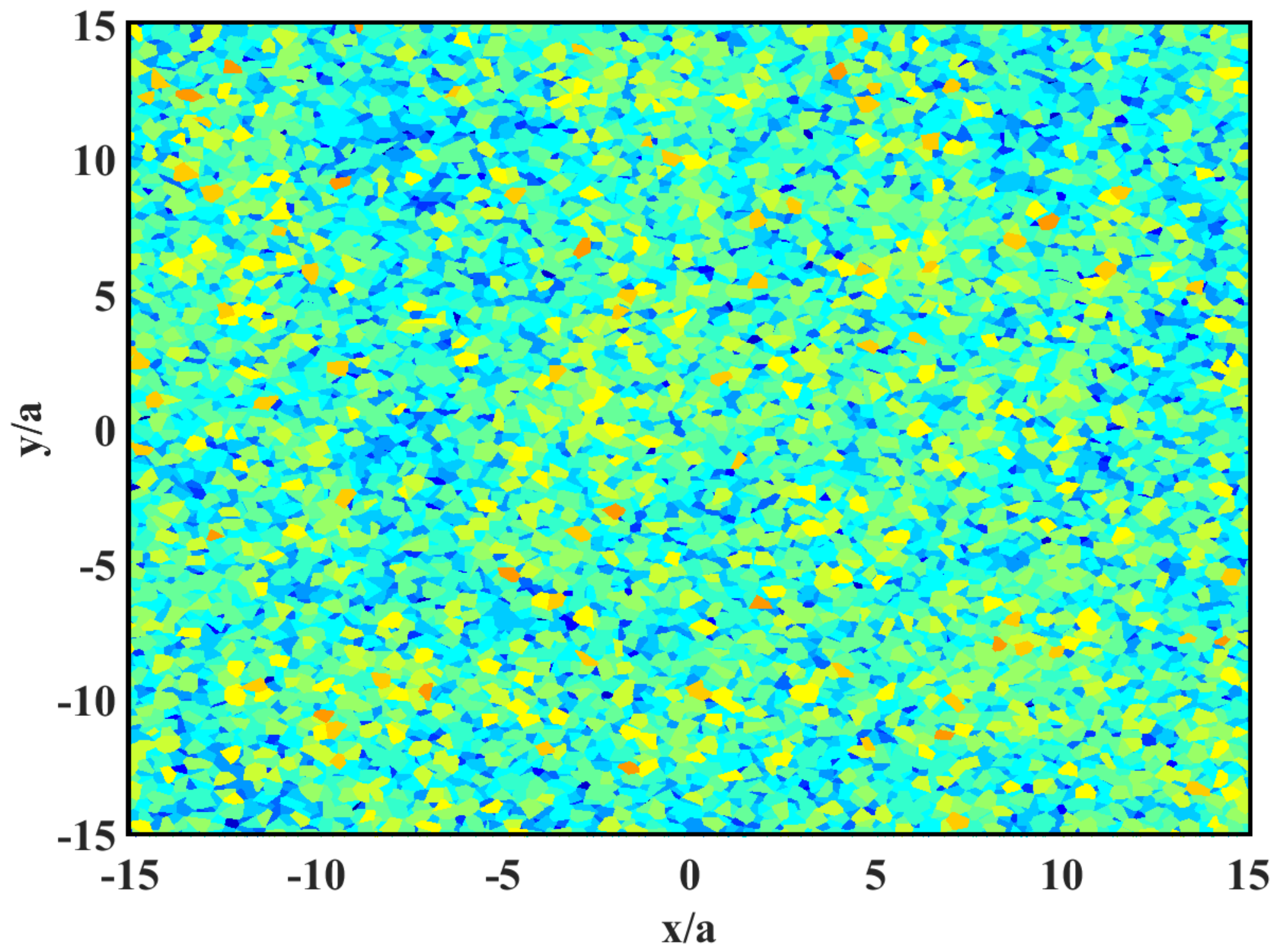}
        \label{fig:melt30x30x10-05.eps}
    \end{subfigure}
        \quad
    \begin{subfigure}{0.45\linewidth}
    \caption{\RaggedRight\textbf{}}
    \includegraphics[scale=0.24]{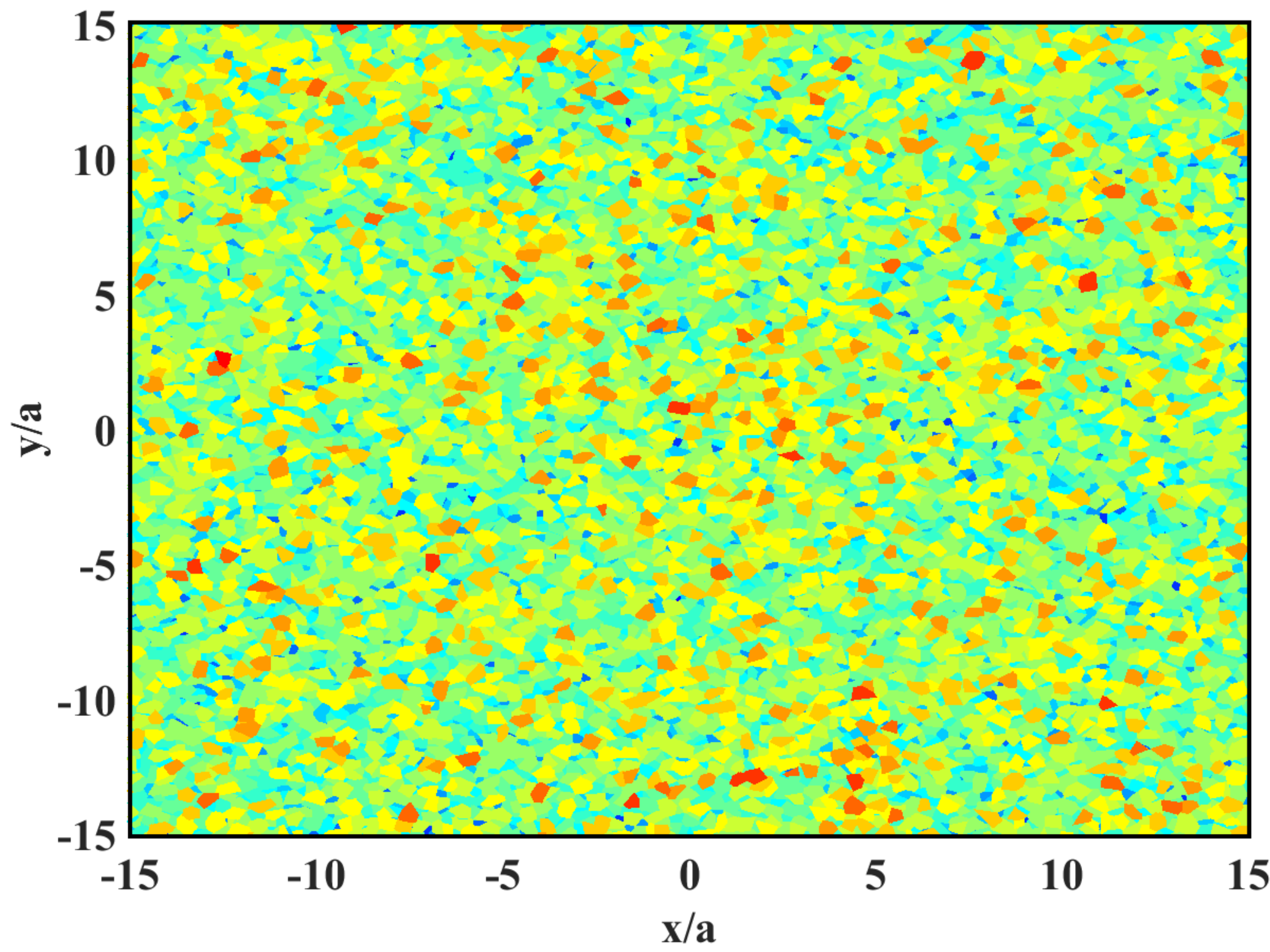}
        \label{fig:Glassbig.eps}
    \end{subfigure}    
    \par\bigskip
    \begin{subfigure}{0.45\linewidth}
    \caption{\RaggedRight\textbf{}}
    \includegraphics[scale=0.24]{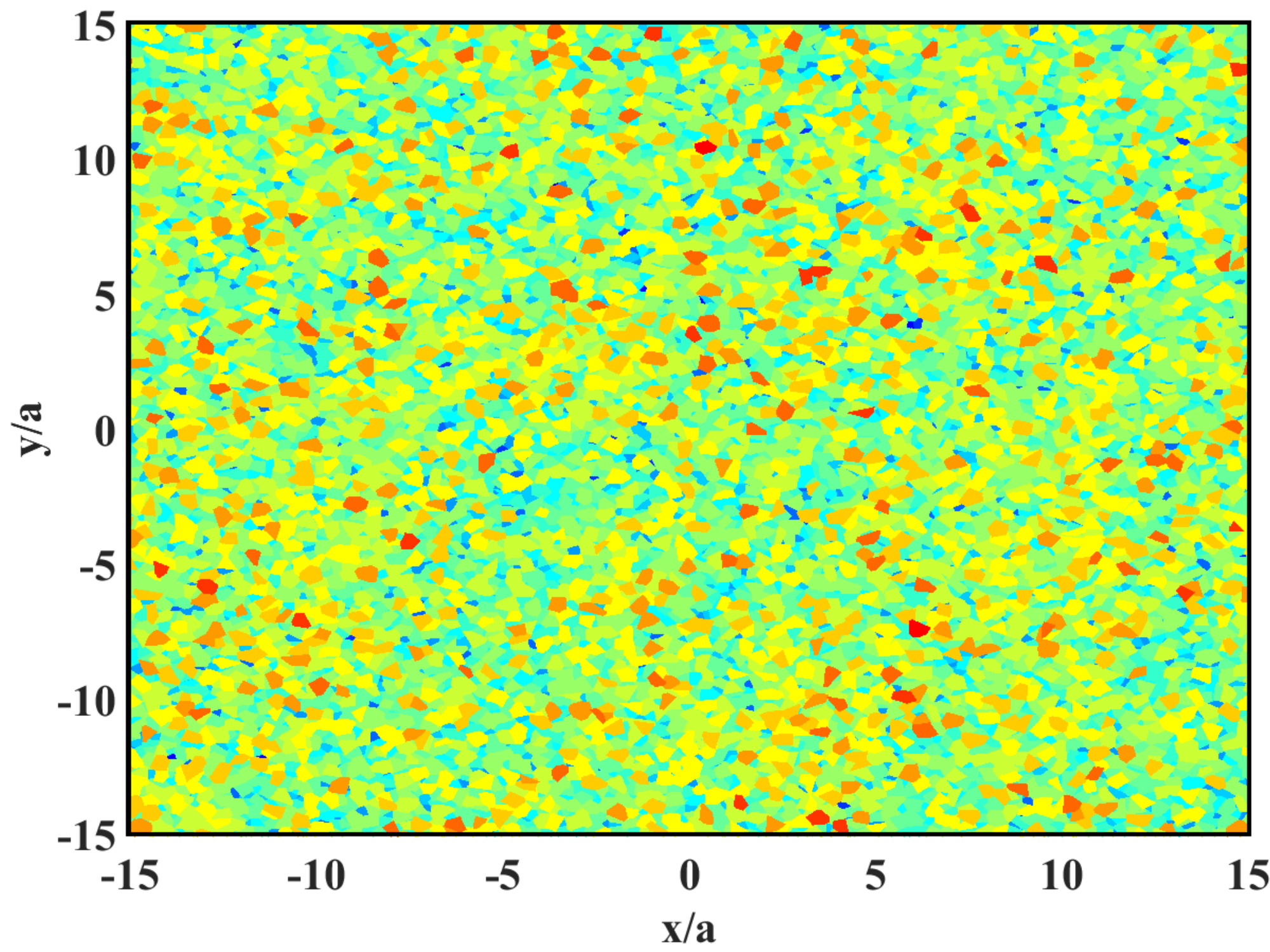}
        \label{fig:5percentbig.eps}
    \end{subfigure}
    \quad
    \begin{subfigure}{0.45\linewidth}
    \caption{\RaggedRight\textbf{}}
    \includegraphics[scale=0.24]{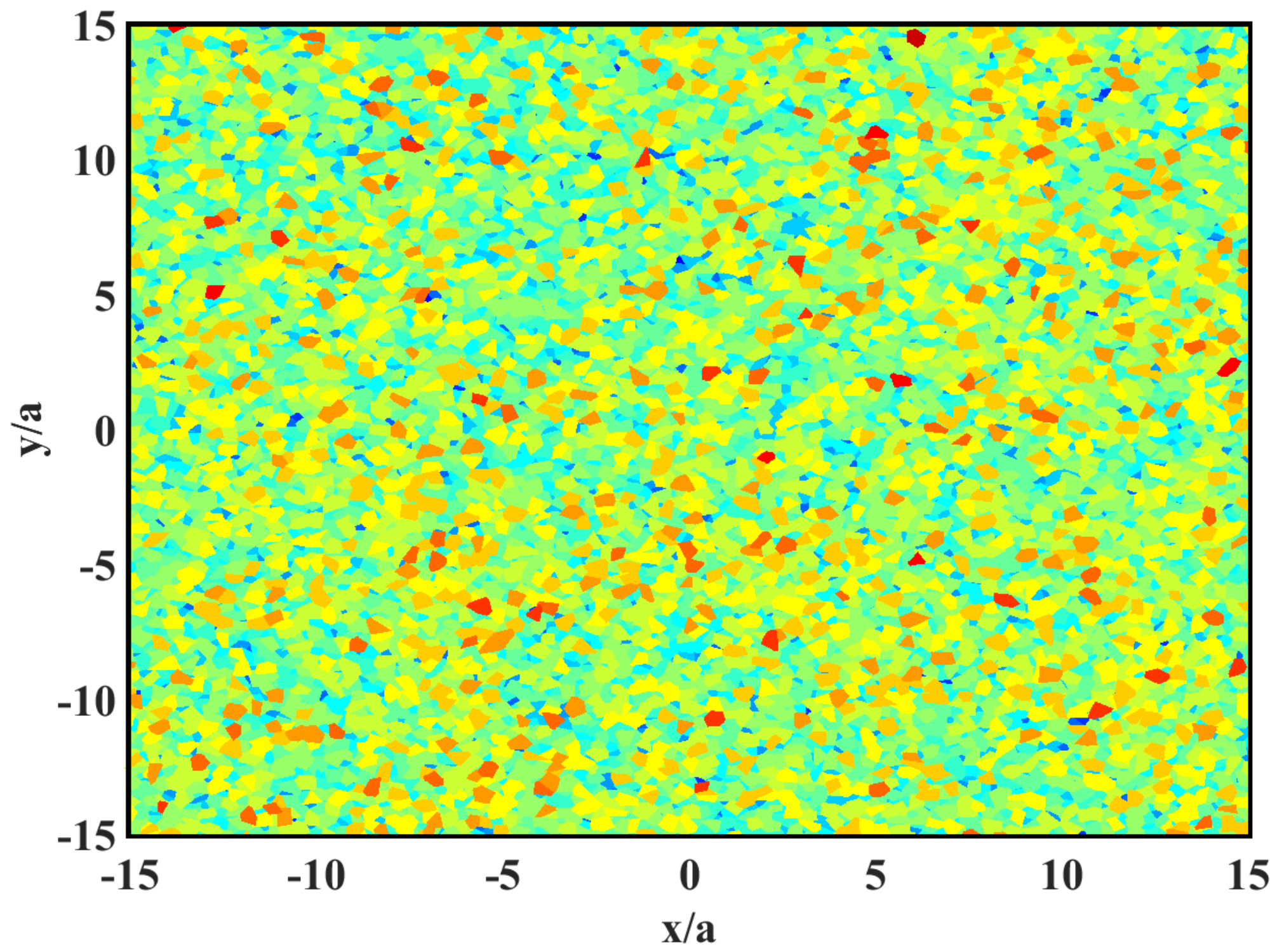}
        \label{fig:10percentbig.eps}
    \end{subfigure}
    \par\bigskip
    \begin{subfigure}{0.45\linewidth}
    \caption{\RaggedRight\textbf{}}
    \includegraphics[scale=0.24]{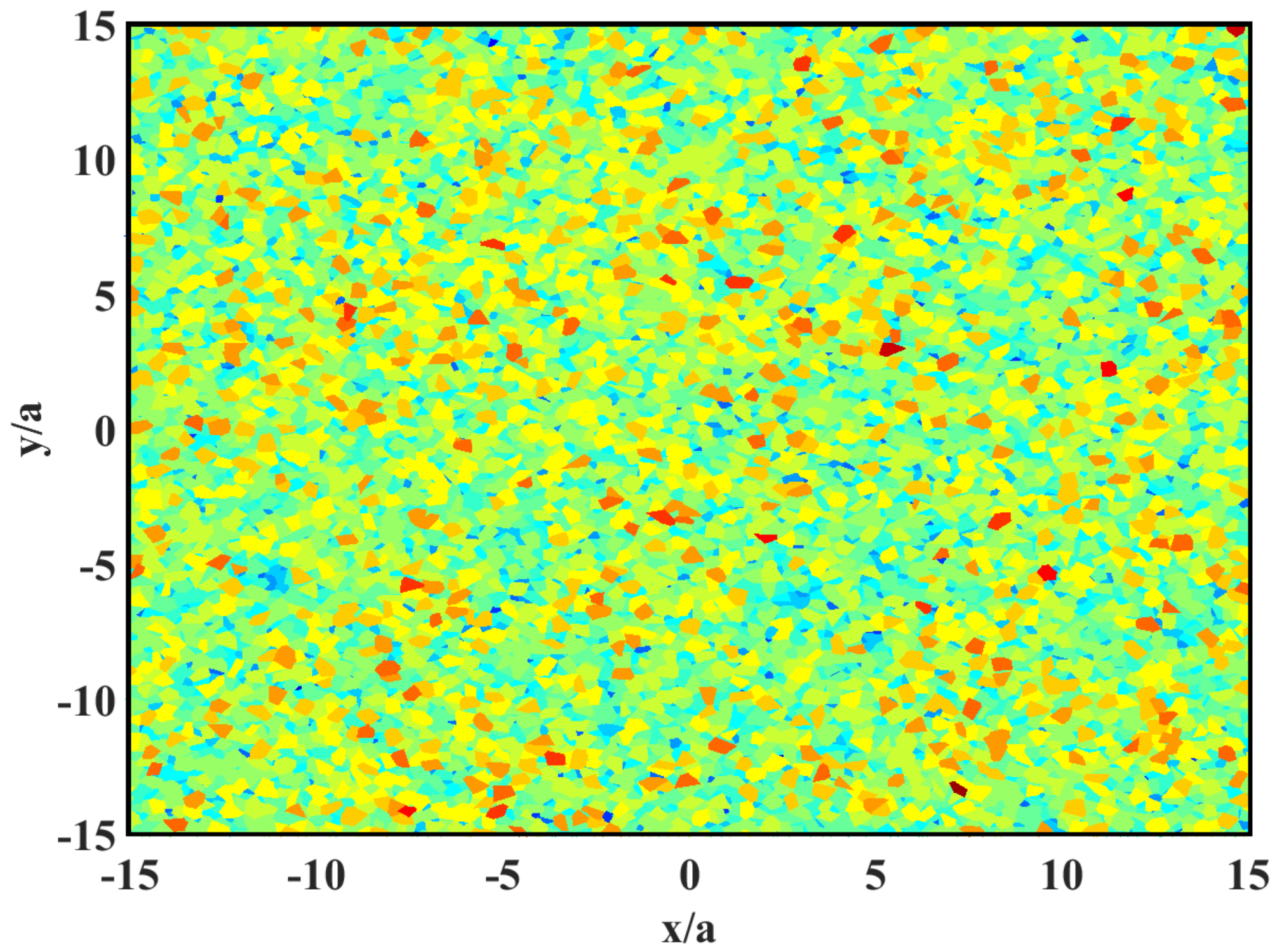}
        \label{fig:15percentbig.eps}
        \end{subfigure}
    \quad
    \begin{subfigure}{0.45\linewidth}
    \caption{\RaggedRight\textbf{}}
    \includegraphics[scale=0.24]{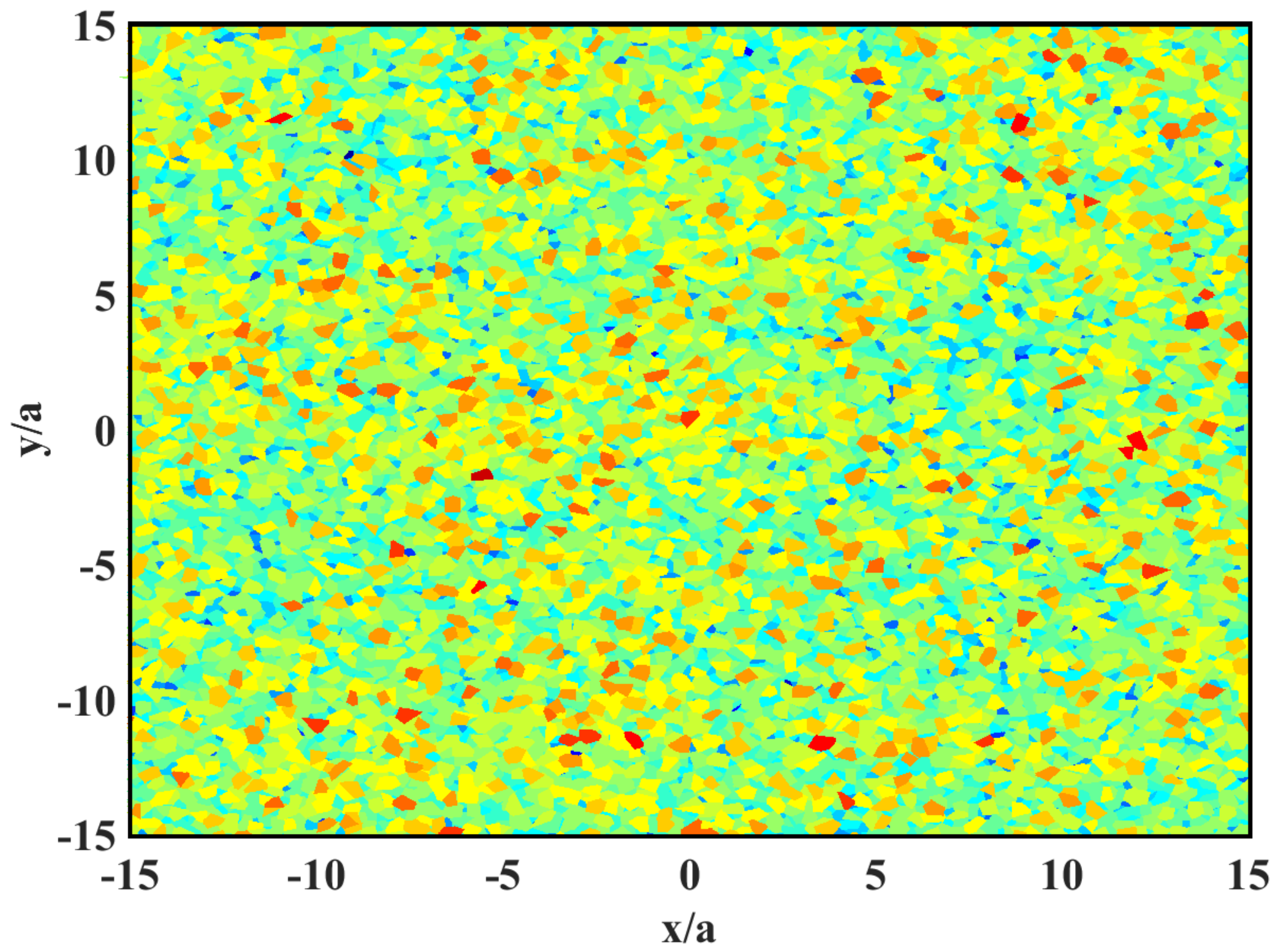}
        \label{fig:20percentbig.eps}
    \end{subfigure}
    \par\bigskip
    \begin{subfigure}{0.45\linewidth}
    \caption{\RaggedRight\textbf{}}
    \includegraphics[scale=0.24]{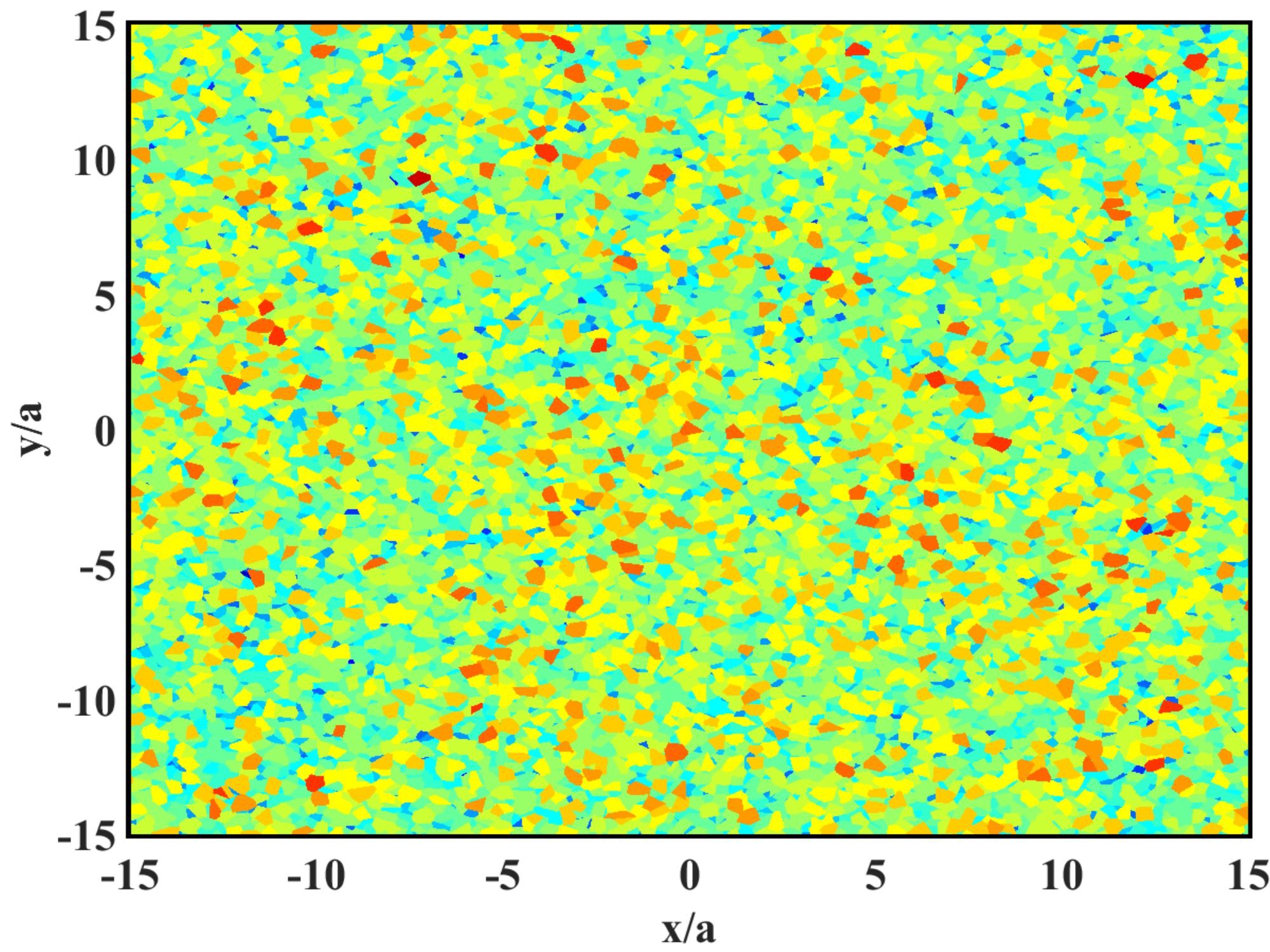}
        \label{fig:25percentbig.eps}
    \end{subfigure}
    \quad
    \begin{subfigure}{0.45\linewidth}
    \caption{\RaggedRight\textbf{}}
    \includegraphics[scale=0.24]{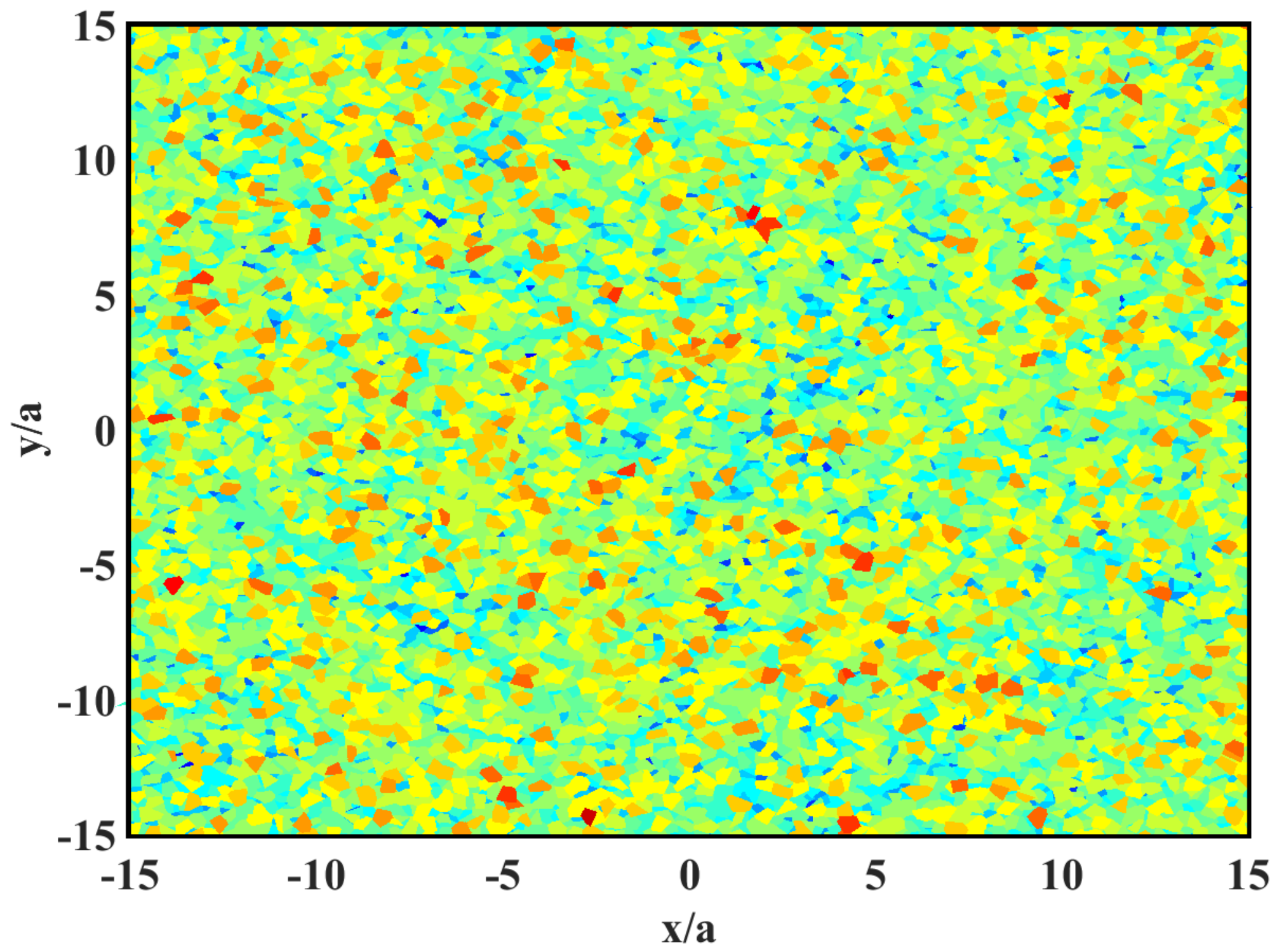}
     \end{subfigure}
    \caption{Local number density distribution in the ``broad zone'' of size $30a\times30a\times10a$ near the interface. \textbf{a}, Color bar for the local number
     density. \textbf{b-i}, Local number density distribution in the melt at $T$=$0.5u_o/K_b$, glass at $T$=$0.3u_o/k_B$, and deformed samples at $5$\%, $10$\%, $15$\%, $20$\% ,$25$\%,\& $30$\% plastic strains, respectively, at $T$=$0.3u_o/K_b$. Local number density is calculated by considering $2.5a$ unit spheres around the beads. \label{fig:broad-zone-density}}
    \end{figure}

Figure ~\ref{fig:broad-zone-density} Distribution of local number density in the ``broad zone'', described in the main letter, for equilibrated melt (at T=$0.5 u_o/k_B$), undeformed glass (at T=$0.3 u_o/k_B$), and deformed glass at 5\%, 10\%, 15\%, 20\%, 25\%, and 30\% plastic strains (at T=$0.3 u_o/k_B$).
The local number density distribution in ``broad  zone'' is similar to what was found in the ``narrow zone'', i.e., equilibrated melts have lower local number density, and in a deforming glass the local number density changes as deformation proceeds; however, there is no indication of increased liquidity in a deforming glass. 

\clearpage
\begin{table}[H]
\centering
\caption{Summary of tracking dilatation and densification events, at an interval of  $0.5$ $\tau$, for chain-ends.}
\begin{tabular}{|l|l|l|l|l|}
\hline
\textbf{Chain-end } & \textbf{Number of }&\textbf{Number of}& \textbf{Number of } & \textbf{ Ration of } \\ 
                 ID                & \textbf{dilatational}   &   \textbf{volume neutral} &  \textbf{compaction} &  \textbf{compaction to}    \\
                   		  	  &	\textbf{events}	   & \textbf{events}               & \textbf{events}         &  \textbf{dilatation}	\\ 
                                    & (\textbf{$\Delta\rho_N<0$}) & (\textbf{$\Delta\rho_N=0$}) & (\textbf{$\Delta\rho_N>0$}) & (\textbf{$\Delta\rho_N<0$/$\Delta\rho_N>0$}) \\ \hline
$29$-$1$ & $1605$ & $710$ & $1586$ & $1.01198$ \\ \hline
$32$-$500$ & $1573$ & $753$ & $1575$ & $0.99873$ \\ \hline
$222$-$1$ & $1562$ & $734$ & $1605$ & $0.973209$ \\ \hline
$315$-$500$ & $1590$ & $725$ & $1586$ & $1.002522$ \\ \hline
$335$-$500$ & $1587$ & $747$ & $1567$ & $1.012763$ \\ \hline
$420$-$1$ & $1595$ & $704$ & $1602$ & $0.99563$ \\ \hline
$447$-$1$ & $1594$ & $722$ & $1585$ & $1.005678$ \\ \hline
$455$-$500$ & $1555$ & $789$ & $1557$ & $0.998715$ \\ \hline
$465$-$500$ & $1606$ & $690$ & $1605$ & $1.000623$ \\ \hline
$504$-$1$ & $1574$ & $717$ & $1610$ & $0.97764$ \\ \hline
$518$-$1$ & $1597$ & $724$ & $1580$ & $1.010759$ \\ \hline
$565$-$500$ & $1546$ & $791$ & $1564$ & $0.988491$ \\ \hline
$570$-$1$ & $1563$ & $775$ & $1563$ & $1$ \\ \hline
$581$-$1$ & $1632$ & $663$ & $1606$ & $1.016189$ \\ \hline
$595$-$1$ & $1566$ & $754$ & $1581$ & $0.990512$ \\ \hline
$647$-$500$ & $1570$ & $723$ & $1608$ & $0.976368$ \\ \hline
$691$-$500$ & $1559$ & $755$ & $1587$ & $0.982357$ \\ \hline
$701$-$500$ & $1583$ & $763$ & $1555$ & $1.018006$ \\ \hline
$744$-$500$ & $1583$ & $761$ & $1557$ & $1.016699$ \\ \hline
$817$-$1$ & $1544$ & $824$ & $1533$ & $1.007175$ \\ \hline
$870$-$500$ & $1583$ & $776$ & $1542$ & $1.026589$ \\ \hline
$937$-$1$ & $1556$ & $772$ & $1573$ & $0.989193$ \\ \hline
$990$-$500$ & $1583$ & $723$ & $1595$ & $0.992476$ \\ \hline
\end{tabular}
    \label{tab:dilatation-densification-data}

\clearpage
\begin{figure}[H]
\centering 
    \begin{subfigure}{0.45\linewidth}
    \includegraphics[scale=0.35]{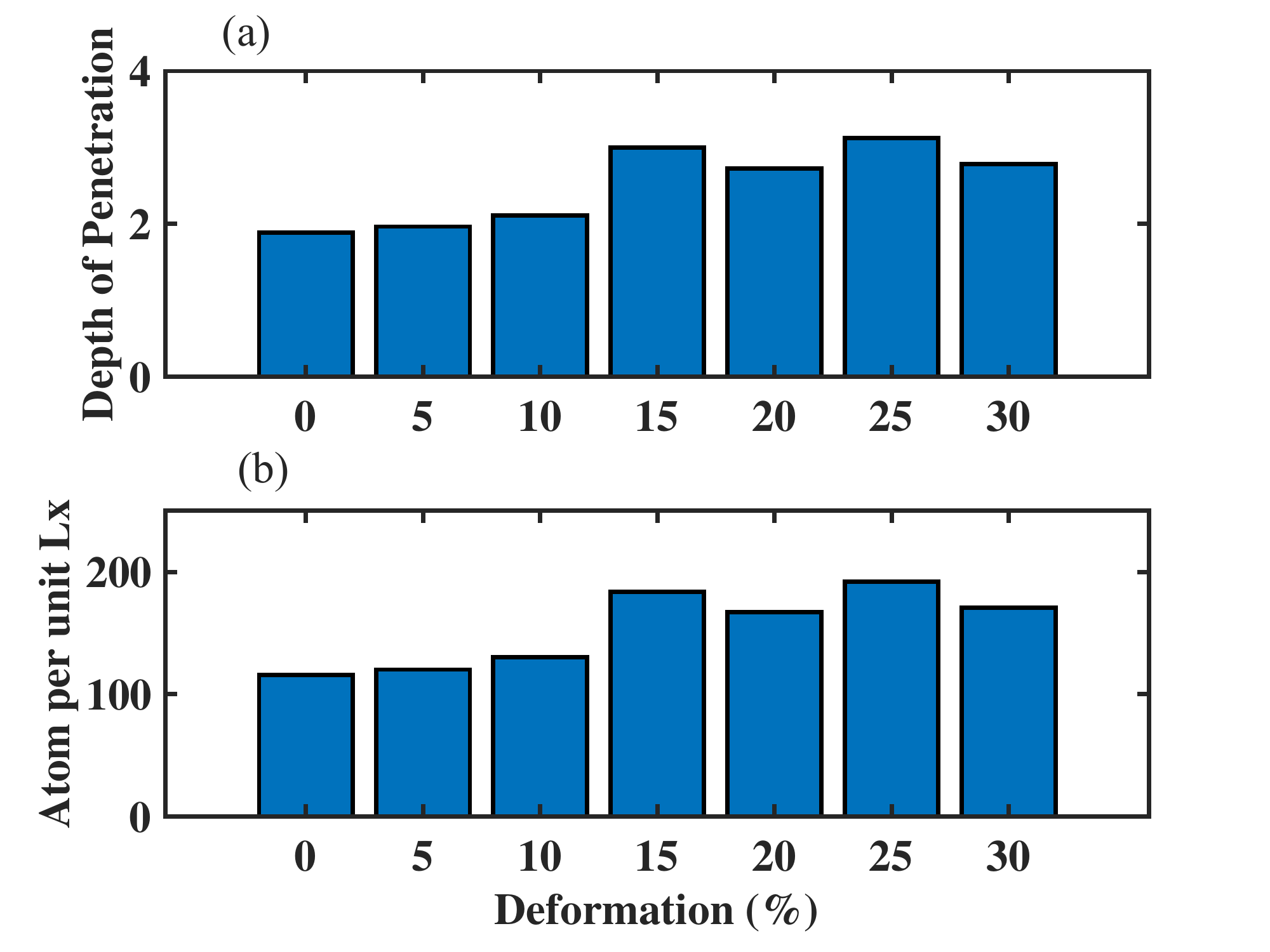}
        \label{fig:Bargraph.eps}
    \end{subfigure}
        \quad
  \caption{\RaggedRight{(a) Depth of penetration or interfacial region with respect to plastic deformation, and (b) number of atoms per unit interfacial width.     \label{fig:depth-of-penetration}}}
    \end{figure}
  \end{table}

Figure ~\ref{fig:depth-of-penetration} shows the growth of interfacial region as plastic strain is imposed. At $0$\% plastic strain the two sample boxes under Van der Waals interaction come closer into molecular proximity, even before plastic deformation is imposed, such that, they create a molecular-scale overlap. We refrain from ``removing'' this overlap from our data.  It is seen that as plastic deformation proceeds, the interfacial region expands (in the $Z$-direction),  
Figure ~\ref{fig:depth-of-penetration}(a) and, new beads per unit interfacial width enter the interfacial region, Figure ~\ref{fig:depth-of-penetration}(b). Both trends are non-monotonic.

\begin{figure}[H]
    \centering
    \includegraphics[scale=0.5]{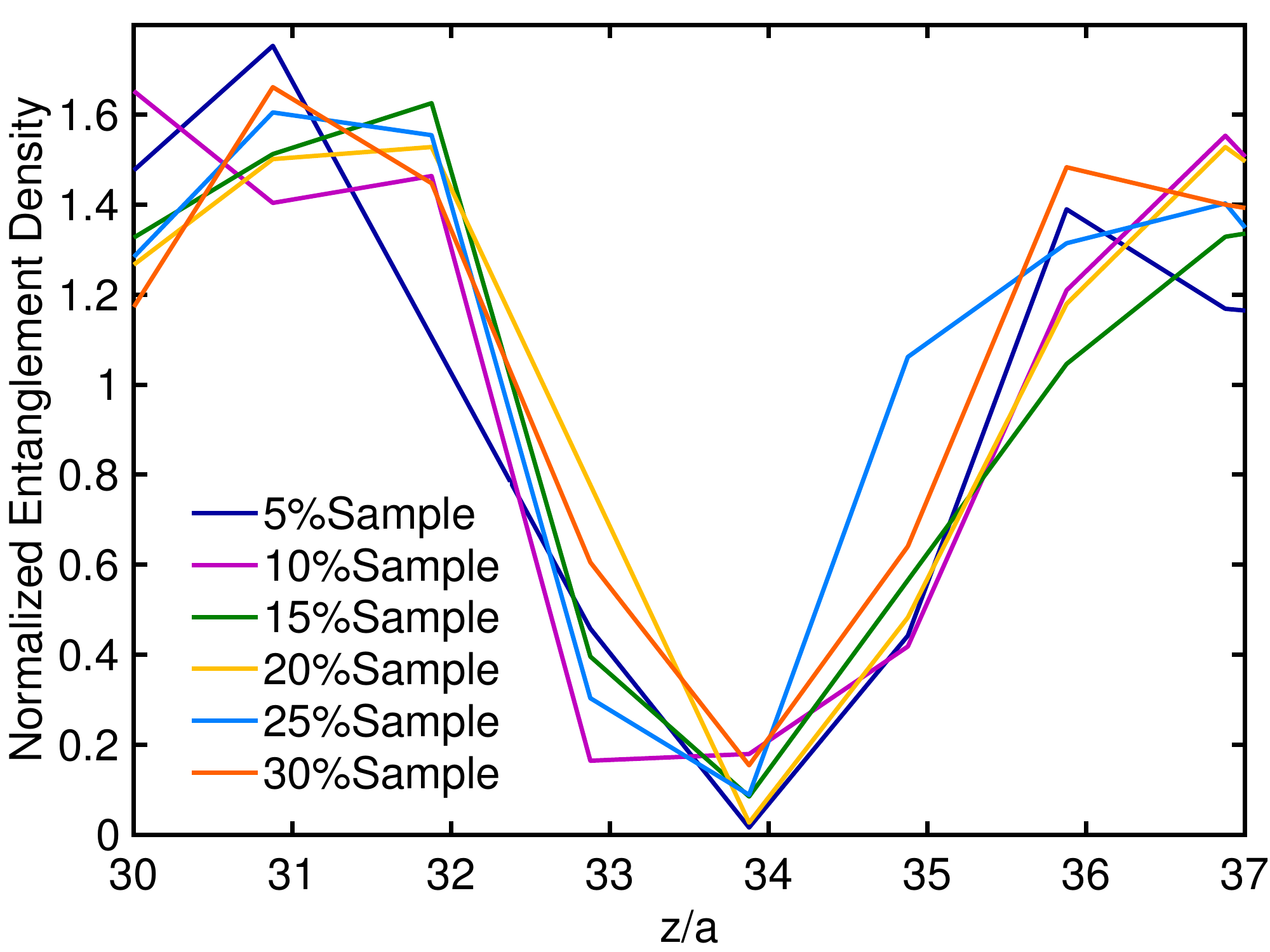}
    \caption{Normalized entanglement density with respect to plastic strain. The bonding interface is centered around $z/a$=$33.5$.}
    \label{fig:entanglement-density}
\end{figure}

We utilized the Z1-code to study the growth of entanglements near the interface (located at $z/a$=$33.5$). 
As compression proceeds in the $Z$-direction, and plane strain conditions are maintained in the $Y$-direction, 
 the sample expands in the $X$-direction, therefore we computed the normalized entanglement density 
 (defined as the total number of entanglements in $a$ units thick layer along the Z-direction, divided by the 
the length of the bonded interface in the $X$-direction). The plot of the normalized entanglement density 
with respect to the $Z$-dimension, is shown in Figure ~\ref{fig:entanglement-density}. We note
that as plastic strain is imposed, the normalized entanglement density first increases from 5\% to 10\% plastic strain, then decreases for 
15\%, and 20\% plastic strains, and then again increases at 25\%, and 30\% plastic strains. This non-monotonic behavior is 
consistent with the non-monotonic trends of $W_f$ with respect to plastic strain, both, in our molecular simulations and experiments. 

\begin{figure}[H]
    \centering
    \includegraphics[scale=0.43]{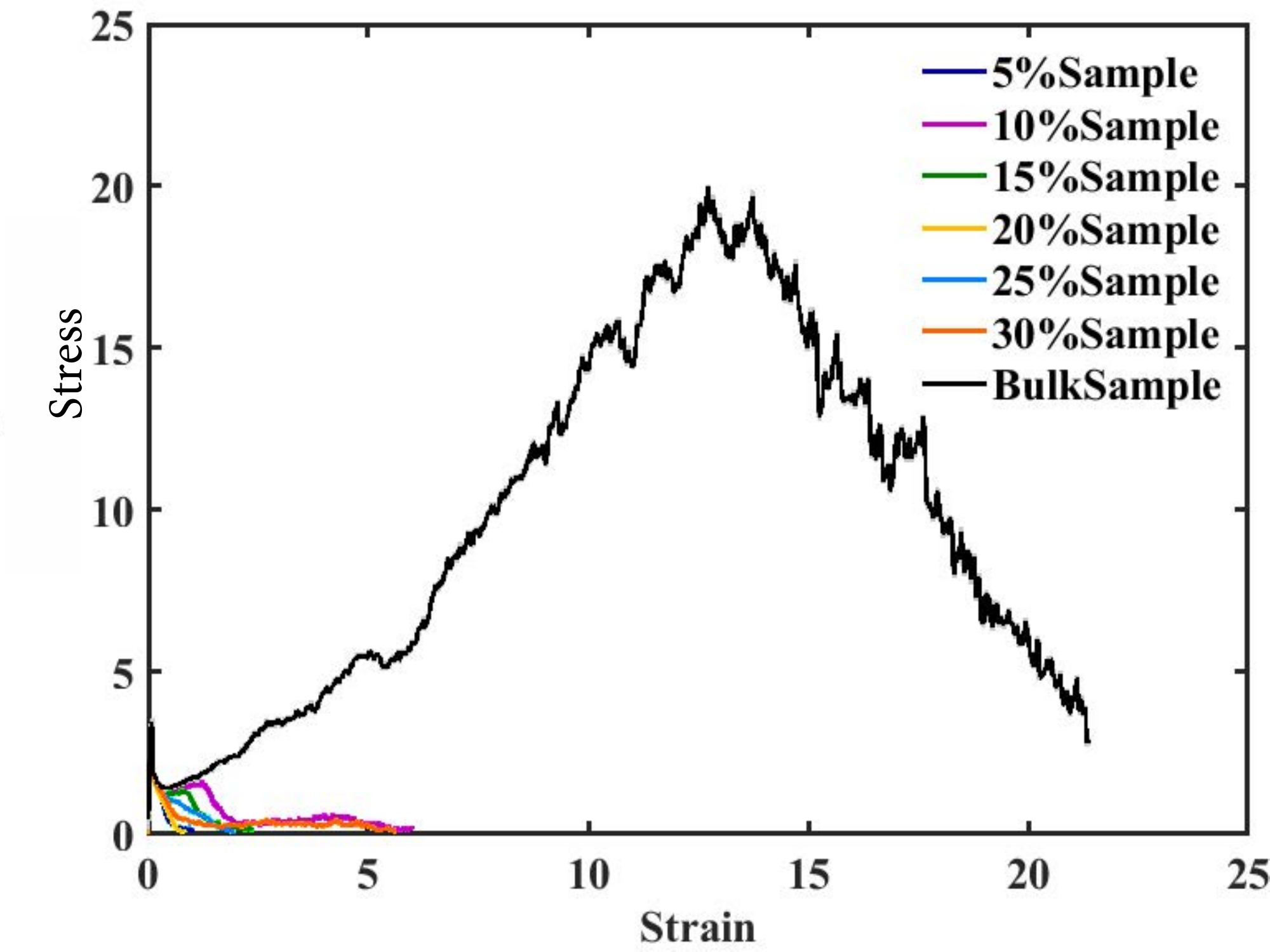}
    \caption{Comparison of bonding strengths with bulk tensile test for samples with $N=500$. \label{weak-bonding}}
\end{figure}

\begin{figure}[H]
    \centering
    \includegraphics[scale=1.5]{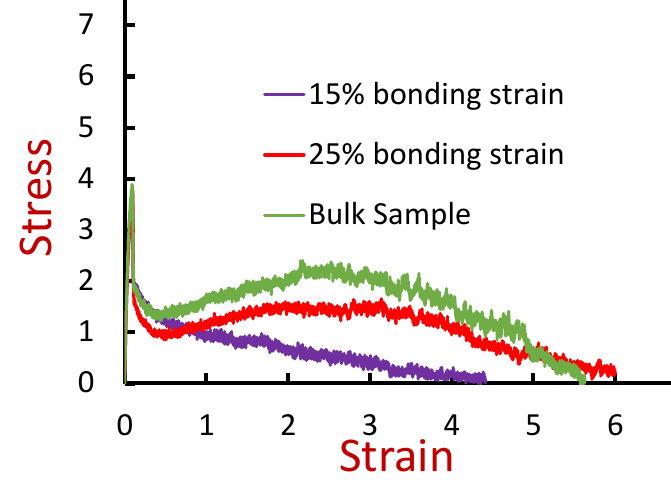}
    \caption{Strong bonding through deformation for polymer samples with $N=100$.}
    \label{fig:strong-bonding}
\end{figure}

For the samples considered in this study so far, bonding experiments at $T=0.3 u_o/k_B$ revealed that through deformation, ductile 
strengths only up to 11\% of the bulk strength were obtained, as shown in Figure ~\ref{weak-bonding}. We recollect that the bonding temperature 
$T=0.3 u_o/k_B$ is significantly lower than the bulk glass-transition temperature ($T_g^b=0.445 u_o/k_B$), and as such the relatively weak 
bonding noted is not surprising. To quickly verify that strengths comparable to that 
of bulk can be rapidly obtained via deformation-induced bonding, we carried out bonding experiments 
on a polymer system at $T=0.44 u_o/k_B$ with $N=100$, and bulk T$_g=0.443 u_o/k_B$, and compared them with the bulk strength of the same polymer, Figure ~\ref{fig:strong-bonding}. This choice of material and processing condition was motivated by the hypothesis that: (i) lowering molecular weight increases
 the number of chain-ends on the free surface of the sample, and we have already seen that chain-ends play a critical role in interpenetration 
 and formation of entanglements, which yields stronger bonding, and (ii) increasing the temperature will make the polymer matrix more ductile,
 and deformation will amplify the already existing high mobility to yield quick bonding.   
Clearly, we are able to obtain bonding strengths approaching the bulk strength, within 80\%. These results provide sufficient confidence that 
optimization in processing parameters (strain rates, and temperatures), and even material conditions, can yield bulk bonding strengths rather quickly via 
deformation, and deformation-acceleration in thermoplastics has opened a new mechanistic pathway unknown to the scientists and practitioners, thus far.  
It should be noted that absolute strengths of sample with $N=100$ has reduced compared to $N=500$, however, the fact that we are able to 
start approaching bulk-like strengths is a remarkable result in its own right. We speculate that bulk strengths can also be achieved for higher molecular weight 
glasses (like $N=500$) via deformation quickly (compared to long-time interdiffusion), and will require detailed optimization of processing conditions, or even 
alterations in the loading program. We defer such interesting and potentially transformative explorations for the future.

\section*{Acknowledgement}

N. P. acknowledges David M. Parks, especially, for pointing out in the past that dilatations during plastic deformation could play a crucial role in deformation-induced bonding. N. P. also thanks Gregory C. Rutledge for providing key references on the topics of molecular simulations, and other helpful discussions during the course of this study. We thank Martin Kroger for providing us the Z1-code.

\end{document}